\documentclass[manuscript,screen]{acmart}
\AtBeginDocument{%
  }

\setcopyright{none} 
\copyrightyear{2026}
\acmYear{2026}
\acmDOI{}
\acmConference[]{}{}{}
\acmISBN{978-1-4503-XXXX-X/2018/06}

\usepackage{subcaption}
\usepackage{multirow}

\begin{document}

\title[Simulating Word Suggestion Usage]{Simulating Word Suggestion Usage in Mobile Typing to Guide Intelligent Text Entry Design}

\author{Yang Li}
\orcid{0009-0007-3697-4981}
\affiliation{%
  \institution{Saarland Informatics Campus, Saarland University}
  \city{Saarbrücken}
  \country{Germany}
}
\email{yang.li@uni-saarland.de}

\author{Anna Maria Feit}
\orcid{0000-0003-4168-6099}
\affiliation{%
  \institution{Saarland Informatics Campus, Saarland University}
  \city{Saarbrücken}
  \country{Germany}
}
\email{feit@cs.uni-saarland.de}

\renewcommand{\shortauthors}{Li and Feit}

\begin{abstract}
Intelligent text entry (ITE) methods, such as word suggestions, are widely used in mobile typing, yet improving ITE systems is challenging because the cognitive mechanisms behind suggestion use remain poorly understood, and evaluating new systems often requires long-term user studies to account for behavioral adaptation. We present WSTypist, a reinforcement learning-based model that simulates how typists integrate word suggestions into typing. We extend recent hierarchical control models of typing, by identifying and implementing important cognitive mechanisms that underlie the high-level decision-making for integrating word suggestions into manual typing: considering orthographic processes, assessing efficiency gains, and including personal preference on AI support. Our evaluations show that WSTypist simulates diverse human-like suggestion-use strategies, reproduces individual differences, and generalizes across different systems. Importantly, we demonstrate on four design cases how a computational rationality model can be used to inform what-if analyses during the design process, by simulating how users might adapt to changes in the UI or in the algorithmic support, reducing the need for long-term user studies.
\end{abstract}




\maketitle

\section{Introduction}
\label{sec:intro}

Modern touchscreen devices often include intelligent text entry systems (ITEs), such as word suggestions to reduce typing effort and improve efficiency~\cite{palin2019people}. Prior work~\cite{alharbi2020effects, lehmann2023typing, fowler2015effects, kristensson2021design, quinn2016cost} shows that suggestions can reduce manual keystrokes and are often perceived as helpful, but they also introduce additional visual and cognitive costs. Already without suggestions, mobile typists must split their visual attention between the keyboard and input field~\cite{jiang2020we, jokinen2021touchscreen} to guide finger movements and proofread text. The suggestion list adds another focus of attention, creating a tension between potential efficiency gains and the interruptions caused by shifting gaze, checking suggestions, and selecting words.

Empirical research shows substantial individual differences in suggestion use. Some users rely on suggestions for completion, error correction, or anticipating upcoming words, while others, especially faster typists, largely ignore them~\cite{lehmann2023typing, palin2019people}. Those who make use of suggestions were found to check the suggestion list much more often than they select from it or do not pick a word even if it is correctly suggested~\cite{2025suggestion}. This indicates that usage depends not only on system quality but also on individual policies for allocating visual and cognitive resources. Understanding how such policies emerge and respond to alternative system designs is important for advancing intelligent text entry. However, collecting empirical data on these policies is costly, particularly when eye-tracking is involved~\cite{hutton2019eye}, and users often require time to adapt to new keyboard designs~\cite{jokinen2017modelling}. Computational simulation models~\cite{murray2022simulation} offer a practical alternative by providing a flexible framework for generating user behavior. Yet prior models of mobile typing either neglect the cognitive mechanisms underlying word suggestion use~\cite{jokinen2021touchscreen, shi2024crtypist}, including the linguistic processing of to-be-written text,  or rely on manually specified abstractions of gaze and motor behavior rather than modeling them as part of the decision process~\cite{kristensson2021design}.

To address this challenge, we propose WSTypist, a simulation model that predicts how users balance the visual and cognitive demands of suggestion use with manual typing and proofreading. WSTypist builds on hierarchical control models of typing within the computational rationality framework~\cite{jokinen2021touchscreen, shi2024crtypist}. To capture the high-level decisions that determine when and how typists attend to, evaluate, and act on word suggestions during normal typing, WSTypist implements three mechanisms that enable the supervisory agent to learn when suggestions are beneficial and when manual typing is preferable, reflecting the diverse ways human typists balance effort, accuracy, and preference~\cite{2025suggestion, lehmann2023typing}: (1) \textit{Orthographic processing}, modeling how spelling processes and linguistic knowledge, together with linguistic properties (e.g., syllable boundaries, lexical frequency), influence suggestion-use behavior; (2) \textit{Efficiency assessment}, evaluating potential keystroke savings and proofreading demands under working memory constraints; and (3) \textit{Personal preference}, capturing individual tendencies such as offloading cognitive effort, satisfaction when system behavior matches expectations, or a preference for uninterrupted manual typing.

To train and evaluate the model, we introduce a custom suggestion system and empirically grounded benchmarks that capture not only typing speed and error rates but also checking behavior and suggestion usage patterns. Our evaluation shows that WSTypist reproduces human-like typing across multiple metrics (e.g., suggestion usage, gaze allocation, typing speed) and user groups (e.g., very fast or slow typists), generalizes to systems without suggestions or auto-correction, and replicates diverse usage strategies such as word completion, typo correction, capitalization, and selecting a similar word before completing it manually~\cite{lehmann2023typing}.

We then demonstrate how computational rationality models can support system design by simulating how users adapt their behavior under alternative design choices. In four use cases, we show how WSTypist enables what-if analyses regarding behavioral adaptation: (1) examining how suggestion accuracy influences typing behavior, showing that higher accuracy increases suggestion use but may also lead to overreliance and more failures, with accuracies around 60–70\% offering the best trade-off; (2) testing a strategy that prioritizes longer words in the suggestion algorithm, which improved typing efficiency in simulations; (3) exploring support of personalized suggestion strategies, finding that prioritizing capitalized words could benefit users who frequently type proper nouns; and (4) evaluating an alternative UI design that places the top suggestion directly in the input field and makes it accessible via a keyboard shortcut, which leads to improved suggestion efficiency and overall typing performance in simulations.

In summary, WSTypist advances existing simulation models of mobile typing behavior by modeling the orthographic processes involved in typing and how they influence suggestion usage in addition to personal preference and overall efficiency assessment. Our results demonstrate that WSTypist achieves ecological validity in reproducing human typing behavior with and without word suggestions and show how computational rationality models can directly inform the evaluation and design of intelligent text entry systems. We open-source the model and the suggestion system, making both readily available to the research community for evaluation and system design.\footnote{\url{https://osf.io/wahkr/overview?view_only=f2c86362be234026a5eee57bca04b500}}.

\section{Related Work}
\label{sec:related}

We review prior work on typing behavior with word suggestions, suggestion system design, and simulation-based modeling of text entry. We focus on how users allocate visual attention and interact with suggestion interfaces, and identify gaps in the joint modeling of gaze, motor actions, and decision-making when using word suggestions.

\subsection{Understanding typing behavior with word suggestions and suggestion system design}
To support efficient text entry on touchscreen devices, numerous Intelligent Text Entry (ITE) methods have been developed~\cite{kristensson2009five}. Gesture input~\cite{leiva2021we, kristensson2004}, for instance, allows users to slide their fingers across the keyboard, while auto-correction automatically fixes typos or applies adjustments such as capitalization~\cite{banovic2019limits, alharbi2020effects}. Another widely adopted approach is word suggestions, which was originally developed for augmentative and alternative communication devices~\cite{darragh1990reactive, garay1997intelligent} to partially automate text input and thus increase communication speed and ease, particularly for people with physical disabilities. Today, word suggestions are a standard feature of mobile keyboards~\cite{leino2024mobile, palin2019people}, typically displayed above the keyboard and updated in real time based on the user’s ongoing input.

While suggestion lists can reduce keystrokes, prior studies show they may also introduce cognitive overhead~\cite{alharbi2020effects, lehmann2023typing, fowler2015effects, kristensson2021design, palin2019people, quinn2016cost, anastaseni2025smartphones, roy2025word} due to additional visual attention shifts among the keyboard, input field, and suggestion list, as well as the cognitive effort required to evaluate options~\cite{quinn2016cost, lehmann2023typing, 2025suggestion, koester1996effect}. Despite these costs, users often perceive suggestions as easier and less effortful~\cite{palin2019people, quinn2016cost}. To better understand usage patterns, Lehmann et al.~\cite{lehmann2023typing} identified eight distinct use cases, ranging from Completion and Correction to adjustments like Capitalization and Contraction. More recently, gaze-based analyses~\cite{2025suggestion} revealed inefficiencies such as frequent failed checks, unnecessary manual typing despite available suggestions, and delays due to attentional shifts. Recent psycholinguistic studies further show that suggestion usage is also influenced by orthographic processing during typing, such as users tending to select suggestions more often at syllable boundaries~\cite{kandel2023written, anastaseni2025smartphones}.

Another strand of research focuses on the design of suggestion systems. Interface variations include displaying predictions inline with the input field (e.g., Gmail Smart Compose~\cite{chen2019gmail}) or across device types~\cite{roy2021typing}. The choice of ranking strategy also matters: early systems relied on statistical models, while recent work leverages neural and federated approaches for more personalized and context-aware predictions~\cite{kristensson2018statistical, ghosh2017neuralnetworkstextcorrection, garay2006text, hard2018federated, yu2018device}. These advances highlight key design trade-offs between efficiency, error risk, and cognitive effort, with users often preferring designs they perceive as less demanding even if objective efficiency gains are small~\cite{palin2019people, quinn2016cost}.

However, most prior evaluations focus on isolated metrics such as typing speed or prediction accuracy, overlooking the combined effects of visual attention, cognitive effort, and motor execution. By jointly modeling gaze behavior and finger movements, our approach provides a more integrated account of how users interact with suggestion systems. This enables a more comprehensive evaluation of design trade-offs and supports the development of interfaces that better balance efficiency, usability, and user experience in real-world text entry.

\subsection{Simulation-based modeling of typing}
Simulation and model-based evaluation have long supported HCI by providing predictive accounts of user behavior and reducing reliance on costly user studies (see~\cite{murray2022simulation} for a recent overview). In text entry research, such models have been used for decades to simulate user input, evaluate system performance, explore the effects of design changes under controlled conditions, and optimize system designs~\cite{feit2018, hetzel2021, kristensson2021design, jokinen2017modelling, OPTI}. In the context of word suggestion usage, early work by Koester and Levine~\cite{koester1998model, koester1997keystroke, koester2002modeling} modeled interaction with word prediction systems using keystroke-level models that incorporated motor actions along with simplified perceptual and cognitive processes involved in typing and prediction selection. Their simulations showed that the effectiveness of word prediction is highly sensitive to factors such as prediction accuracy, list organization, and user decision costs, challenging the assumption that prediction universally improves typing performance. However, like other rule-based cognitive models such as ACT-R~\cite{anderson2004integrated} and EPIC~\cite{kieras1997overview}, these approaches rely on manually specified heuristics for decision-making and attention allocation.

More recent models, in contrast, are developed within the computational rationality framework, grounded in supervisory optimal control~\cite{oulasvirta2022, chandramouli2024workflow}. This framework allows agents to learn strategies for allocating attention and planning typing actions under cognitive and environmental constraints. Previous models have focused on predicting eye and finger movements during typing, offering insights into proofreading and error correction behavior~\cite{jokinen2021touchscreen}. Subsequent extensions improved generalization by incorporating pixel-level visual input, expanding benchmarking capabilities~\cite{shi2024crtypist}, and simulating different error sources~\cite{shi2025simulating}. However, these models do not account for intelligent text entry methods such as word suggestions, limiting their applicability to everyday mobile typing. 

Building on this foundation, we introduce a simulation model that extends computational rationality with cognitive mechanisms to capture trade-offs in using word suggestions. The model accounts for both motor processes and higher-level decisions about integrating suggestions, and enables efficient what-if analyses to evaluate design trade-offs and inform more effective intelligent text entry systems.

\section{Simulating Word Suggestion Usage}
\label{sec:methods}

The goal of our work is to facilitate the design of ITE methods by developing a simulation model that allows designers to assess how users adapt their typing behavior and specifically their use of word suggestions (e.g., how often they select them or which strategies they employ~\cite{lehmann2023typing}) and their checking behavior (e.g., when they start checking the suggestion list or how many failed checks occur~\cite{2025suggestion}) in response to changes in system design, such as the suggestion algorithm or the user interface. To achieve this, we build on recent simulation models of mobile typing that have been developed within the computational rationality framework~\cite{oulasvirta2022}, in which typing is formulated as a bounded optimality problem in a Partially Observable Markov Decision Process (POMDP)~\cite{kaelbling1998planning}. Within this framework, we model typing with word suggestions as a hierarchical supervisory control problem in which a supervisory agent learns, via reinforcement learning (RL), an optimal policy for distributing visual attention across the keyboard, input field, and word suggestion list to enable effective checking and selection of suggestions while managing the cognitive and motor demands of manual typing and proofreading. Specifically, we build on a previously proposed architecture~\cite{jokinen2021touchscreen, shi2024crtypist, shi2025simulating} where at each time step, the agent gathers partial observations through eye movements, processes them via a working memory component with limited cognitive capacity and noise, and selects subsequent gaze and finger actions. However, existing models do not account for the cognitive mechanisms underlying suggestion usage, most crucially, the costly and error-prone internal processes that retrieve the spelling of a word at a syllable level~\cite{kandel2023written}. These influence word suggestion usage~\cite{anastaseni2025smartphones}, but also regular typing~\cite{pinet2016typing}. 
In the following, we first describe the mechanisms that influence word suggestion usage, and then detail their implementation.

\subsection{Mechanisms underlying suggestion usage}
\begin{figure*}[t]
  \centering
  \includegraphics[width=0.68\linewidth]{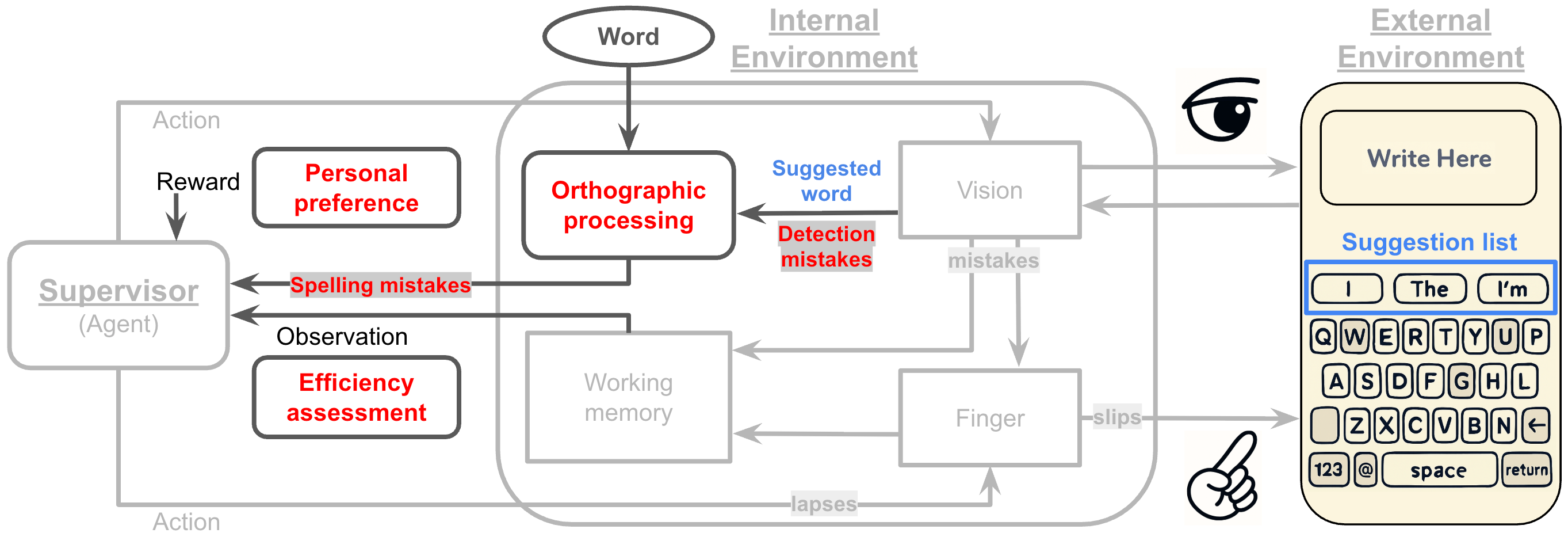}
  \caption{Overall structure of the WSTypist model. We extend an existing hierarchical architecture to simulate suggestion use in mobile typing, where a Supervisor agent learns to optimally control visual attention and finger movements given noisy observations due to working memory constraints (original architecture in gray~\cite{shi2025simulating}). We incorporate key cognitive mechanisms that allow the agent to evaluate the utility of word suggestions, including error-prone orthographic processing for obtaining the to-be-typed characters, assessing the efficiency of word suggestions from noisy observations, and personal preferences that influence the users' perceived reward from a suggestion.}
  \Description{The figure presents a hierarchical model of mobile typing. A central Supervisor agent interacts with an Internal environment containing modules for orthographic processing, working memory, efficiency assessment, and personal preference. The agent receives observations such as spelling errors and system states, and produces gaze and finger actions that influence typing behavior. The Internal environment exchanges information with the External environment, representing a smartphone interface, including a text input area, a suggestion list, and a keyboard. Arrows indicate the flow of information, including word input, suggestion detection, typing actions, and feedback such as errors and rewards.}
  \label{fig:model}
\end{figure*}

Using word suggestions requires weighing their potential benefits against their costs. These decisions vary widely across users~\cite{2025suggestion} and lead to a variety of strategies beyond simple word completion~\cite{lehmann2023typing}, relying on suggestions to insert capitalized letters and apostrophes, or using the suggestion list as a spelling reference~\cite{2025suggestion}. When entering a word, the typist must decide after every character whether to check the suggestion list and possibly select a suggestion or continue typing manually. Selecting a suggestion can reduce effort, for example, by correcting an error that would otherwise require several backspaces or by facilitating or avoiding the spelling of difficult words. However, checking suggestions requires shifting visual attention away from the keyboard or input field, and selection requires moving the finger to the suggestion list, both interrupting the typing flow~\cite{lehmann2023typing}. 
The empirical research discussed in the previous section shows that people's decision to use suggestions depends not only on the external environment, such as the performance of the suggestion algorithm~\cite{roy2021typing} and the interface design~\cite{buschek21, quinn2016cost}, importantly, it also depends on the typist’s cognitive and motor processes, including their orthographic processes~\cite{anastaseni2025smartphones} and individual preferences~\cite{2025suggestion}, which can vary largely across people. 
To capture the varied behavior across users and simulate human-like typing behavior with word suggestions, we extend the recently developed simulation model~\cite{shi2025simulating} in three directions to model the internal cognitive processes that govern suggestion usage. 
These are highlighted in \autoref{fig:model} and described in the following.

\paragraph{Orthographic processing}
Linguistic properties of the target word, such as syllabic structure and the typist’s linguistic skill, particularly orthographic knowledge, influence both typing performance~\cite{feit2016, pinet2016typing} and the use of word suggestions~\cite{2025suggestion, anastaseni2025smartphones}. When typists are uncertain about a word’s spelling, they are more likely to rely on the suggestion list as an external spelling reference~\cite{2025suggestion}. Importantly, research in psycholinguistics showed that orthographic processing operates at the syllable level: encoding upcoming syllables can slow down the typing of the current one~\cite{pinet2016typing}. This mechanism also shapes suggestion usage, with suggestions more frequently selected at syllable boundaries~\cite{anastaseni2025smartphones}. To capture these effects and account for the additional cognitive costs related to retrieving the spelling of words, we introduce an \textbf{Orthographic processing} component into the Internal environment. It models how an abstract representation of a word is encoded into to-be-typed characters at a syllable level. 
This impacts typing in two ways. First, encoding a syllable takes time, and we add a delay in typing speed when additional syllables remain, modeling the ongoing syllable-level spelling processes during writing \cite{kandel2023written, pinet2016typing}.
Second, encoding a word is error-prone, where individual language skills and orthographic uncertainties might lead to wrongly encoded syllables. We model the probability of \textbf{spelling mistakes} based on the frequency of the to-be-typed word and a newly introduced \textbf{Linguistic Knowledge} parameter ($P_K$) that captures individual differences in spelling skills. Importantly, the use of the suggestion list facilitates orthographic processes: when the agent gazes at the suggestion list long enough and identifies the target word without \textbf{detection mistakes}, it acquires the correct spelling, thereby reducing spelling errors and encoding times. 

\paragraph{Efficiency assessment}
One goal of using suggestions is to improve typing efficiency~\cite{2025suggestion}, which is closely tied to keystroke savings~\cite{roy2021typing}. However, recent work~\cite{2025suggestion} shows that selecting a suggestion can cost up to three times more than a regular keystroke, meaning efficiency gains arise only when the remaining keystrokes (including potential error corrections) would take longer than checking and selecting from the suggestion list. To capture this trade-off, we extend the observation space so the supervisory agent can decide whether to consider suggestions both at the word level and after each keystroke. At the word level, information such as target word \textbf{Length} and \textbf{Frequency} influences the likelihood of checking and selecting suggestions~\cite{lehmann2023typing, 2025suggestion}, and informs expectations about whether the keyboard is likely to suggest the word. 
At the keystroke level, decisions depend on estimates of the word’s Certainty (confidence in the currently typed content) and Correctness (accuracy of the current input), introduced in prior work~\cite{shi2024crtypist}, but crucially also on knowledge about the \textbf{Completeness} of the word to decide whether selecting a suggestion will provide keystroke savings. We thus extend the Working memory component to estimate how many characters remain in the current word, subject to working memory limitations.

\paragraph{Personal preference}
Typists also differ in their individual preference for using word suggestions~\cite{palin2019people}, independent of typing expertise. Some users rely heavily on suggestions to reduce workload, whereas others prefer manual typing to retain control or avoid interruptions~\cite{2025suggestion}. Such variation may be related to a person's readiness to adopt new technologies~\cite{parasuraman2000}, prior experience with similar systems, or an individual tendency toward cognitive offloading~\cite{risko2016cognitive}. Users are also more likely to continue using a system when their expectations of its behavior are met~\cite{expectation-confirmation-model}. In the context of word suggestions, this may lead users to keep checking the suggestion list until the desired word appears, confirming that the system provides the expected support. 
We model these aspects within the reward mechanism of the reinforcement learning agent. While previous work assumed that typists optimize their behavior with a focus on typing speed and accuracy, we argue that personal preference and satisfaction with the system also shape behavior. To capture this effect, we add a new term to the reward function based on a \textbf{Suggestion Reliance} parameter ($P_S$). This parameter reflects individual preferences for using the suggestion system and models the perceived reward of selecting a suggestion that matches the target word, independent of objective performance gains.

\subsection{The WSTypist model}
We implement these three mechanisms within an existing hierarchical supervisory control architecture~\cite{shi2025simulating}, the most recent one in a series of works that aim to simulate human-like manual typing on mobile devices. An overview is given in~\autoref{fig:model}. In the following, we describe each component of the model in more detail, focusing on our extensions. We refer to the original papers for more details~\cite{shi2024crtypist, shi2025simulating}.

\subsubsection{Internal environment}
\label{sec:internal}
The Internal environment comprises the \textit{Working memory} and \textit{Orthographic processing} components, as well as \textit{Vision} and \textit{Finger} modules that simulate gaze and finger movements to act on and retrieve information from the External environment.

The Working memory~\cite{baddeley2012working} component captures key cognitive limitations necessary for simulating realistic typing behavior. It receives input from the External environment through the Vision and Finger modules, including time cost and the current typing state, and outputs several metrics to the agent: Certainty, Correctness, and \textbf{Completeness}. Certainty and Correctness have been introduced by previous work and derived through noisy proofreading under cognitive limitations controlled by the Memory parameter ($P_M$)~\cite{shi2024crtypist}. \textbf{Completeness} of the typed word is computed by dividing the current input length ($I_L$) by the target word length ($W_L$), weighted by the agent’s Certainty: $Comp = \frac{I_L}{W_L} \times Cer$.

The Orthographic processing~\cite{anastaseni2025smartphones} component simulates spelling process for upcoming syllables in parallel to the ongoing typing by introducing an additional 50~ms time cost for each encoded syllable, in line with the APOMI theory of word writing~\cite{kandel2023written}. We also include the \textbf{Distance} to the next \textbf{Syllable} boundary $D_S$ in the observation space of the agent. We compute the likelihood of spelling mistakes based on the linguistic knowledge parameter and the word frequency: $E_K = 1 - e^{(-W_F \cdot P_K)}$, while also including other error sources introduced by previous work~\cite{shi2025simulating}. 

The Vision module retrieves information from the External environment based on the most recent gaze action and updates Working memory and Orthographic processing with partial information from the gazed-at location. For example, when gazing at the input field, Vision can access the current input content and update the Certainty in Working memory based on the time spent for proofreading (\textit{Mistakes} in prior work~\cite{shi2025simulating}). When gazing at the suggestion list, Vision can observe whether the target word is currently suggested and how it is spelled. If the gaze shifts to other fields, the suggestion status becomes unknown as the suggestion list is updated. When the gaze is on the keyboard, Vision guides typing to reduce pointing errors. Gaze shifts between interface elements contribute an additional time cost of 200~ms, consistent with previous work~\cite{shi2024crtypist, 2025suggestion}. We use the EMMA model~\cite{salvucci2001integrated} to estimate the time required to read a word in the suggestion list and recognize whether it is the target. This means that less frequent words require more time to be read. Thus, vision actions that are too short to correctly recognize the word can lead to \textbf{detection mistakes}. Visual ability is controlled by a Vision parameter ($P_V$), following previous work~\cite{shi2024crtypist}. 

The Finger module executes typing-related actions in the External environment, including typing the next letter, \textbf{picking a suggestion} from the suggestion list, using backspace to correct errors, or performing no action (NoAct). Finger movement speed is output by the agent as a separate action. Agents are trained across a range of finger movement speeds to reflect variations in human typing speed, based on empirical data~\cite{2025suggestion, jiang2020we}.
Typing errors are included to simulate realistic human behavior. We model the same types of errors as in prior work~\cite{shi2025simulating}, including \textit{Slips} (modeled using the Weighted Homographic model~\cite{guiard2015mathematical} and controlled by the gaze guidance and Finger parameter ($P_F$)~\cite{shi2024crtypist}, as well as unintentional double taps and motor commands swapping), \textit{Lapses} (memory failures in determining where to type next, based on the time elapsed since the last proofreading), and \textit{Mistakes} (arising from imperfect proofreading that depends on the time spent).

\subsubsection{Supervisor agent}
The Supervisor agent collects partial observations of the Internal and External environments, which are processed by the Working memory and the Orthographic processing modules, and makes decisions on how to allocate visual and motor resources to act on the External environment. To enable the agent to learn boundedly optimal policies that are similar to empirically observed strategies and adapt to the limitations of the Internal (cognitive) and External (word-related) environment, we provide the Supervisory agent with the following components, allowing it to better evaluate the cost–benefit trade-offs of suggestion use.

\paragraph{Observation space}
The agent receives a partial observation of the full environment state. We adopt the core items from previous architecture~\cite{shi2024crtypist} and extend the space with additional suggestion-related items. Overall, the observation space thus contains the Length and Frequency of the target word ($W_L$, $W_F$), the Distance to the next Syllable boundary ($D_S$), the current gaze position, and whether the target word is in the suggestion list, the outputs of the three Working memory components (Certainty, Correctness, and Completeness), and the cognitive parameters ($P_M$ for Memory, $P_F$ for Finger, $P_V$ for Vision, $P_K$ for Linguistic Knowledge and $P_S$ for Suggestion Reliance).

\paragraph{Action space}
The action space consists of three separate components: one for gaze movement, one for finger movement, and one for typing speed. Similar to the action given by the supervisor agent every 50~ms from the previous architecture~\cite{jokinen2021touchscreen, shi2024crtypist}, our Gaze action determines where attention is on the keyboard, the \textbf{suggestion list}, or the input field. The Finger action, together with typing speed, can type the next character, press backspace, \textbf{pick a suggestion}, or perform a NoAct (no action).

\paragraph{Reward function}
Previous work~\cite{jokinen2021touchscreen, shi2024crtypist, shi2025simulating} always assumed that typists optimize their behavior with a focus on typing speed and accuracy. However, we argue that users also optimize their behavior to meet personal preferences or subjectively perceived benefits when typing with ITE support. Therefore, these factors should be incorporated into the reward function. We adopt the basic structure of the reward function from previous work~\cite{shi2024crtypist, shi2025simulating}, with an additional component capturing personal preference and perceived satisfaction from suggestion use (the third part). The full reward function is defined as follows:

\begin{equation}
    R = \underbrace{(1 - UER^{\beta})}_{\text{accuracy}} 
        - \underbrace{(\gamma \times \tfrac{T}{W_L})}_{\text{speed}} 
        + \underbrace{(P_S \times \mathbb{I}_{\text{Pick}})}_{\text{suggestion use}}
\end{equation}

Here, $UER$ denotes the final uncorrected error rate, computed by comparing the agent’s final input with the target word. The term $\beta$ scales the penalty for typing errors, while $\gamma$ controls the influence of typing speed. The term $\frac{T}{W_L}$ represents the normalized time cost, where $T$ is the total time for the episode and $W_L$ is the length of the target word. $P_S$ is the \textbf{Suggestion Reliance} parameter, reflecting individual differences in reliance on suggestions and perceived satisfaction from selecting the correct word from the suggestion list. $\mathbb{I}_{\text{Pick}}$ is an indicator function that equals 1 if a suggestion is picked and 0 otherwise, providing a suggestion-related bonus. $\beta$, $\gamma$, and $P_S$ are all positive values. The agent receives this reward only at the end of an episode (i.e., after completing the word). We evaluated the influence of $P_S$ on the model's ability to produce human-like behavior for suggestion usage in Appendix~\ref{sec:abl}, showing the necessity of such mechanism components and their alignment with expected behavior.

\begin{table*}[t]
  \setlength{\tabcolsep}{3.8pt}
  \caption{Comparison of model and human performance under 2-Thumb (2-T), 1-Finger (1-F), and averaged across the two conditions (Avg). We report the Mean (SD) value for each group. The modeling results show close alignment with human data across all metrics, with differences within one SD, especially for key metrics (in bold).}
  \sffamily
  \footnotesize
  \begin{tabular}{lllllllllll}
    \toprule
    && \textbf{Picked} & Failed & Start & \textbf{Gaze Sugg} & Gaze Kbd & BPC & UER & \textbf{WPM }& KS \\
    \midrule
    \textbf{2-T}& Human~\cite{2025suggestion} & \textbf{0.13 (0.17)} & 0.36 (0.06) & 0.42 (0.06) & \textbf{0.17 (0.09)} & 0.32 (0.14) & 0.09 (0.06) & 0.03 (0.02) & \textbf{48.66 (14.65)} & 0.11 (0.07) \\
     & Model & \textbf{0.16 (0.09)} & 0.38 (0.05) & 0.36 (0.04) & \textbf{0.14 (0.05)} & 0.39 (0.08) & 0.07 (0.04) & 0.04 (0.02) & \textbf{42.23 (9.24)} & 0.13 (0.07) \\
    \midrule
    \textbf{1-F} & Human~\cite{2025suggestion} & \textbf{0.23 (0.12)} & 0.42 (0.10) & 0.37 (0.08) & \textbf{0.23 (0.10)} & 0.40 (0.14) & 0.07 (0.04) & 0.04 (0.04) & \textbf{36.50 (6.99)} & 0.19 (0.13) \\ 
     & Model & \textbf{0.22 (0.09)} & 0.45 (0.07) & 0.34 (0.04) & \textbf{0.18 (0.06)} & 0.45 (0.09) & 0.06 (0.03) & 0.05 (0.03) & \textbf{34.92 (7.18)} & 0.18 (0.07) \\
    \midrule
    \textbf{Avg} & Human~\cite{2025suggestion} & \textbf{0.18 (0.11)} & 0.38 (0.08) & 0.40 (0.07) & \textbf{0.20 (0.10)} & 0.36 (0.14) & 0.08 (0.05) & 0.04 (0.03) & \textbf{42.65 (13.03)} & 0.14 (0.11) \\
    & Model & \textbf{0.19 (0.08)} & 0.41 (0.05) & 0.35 (0.04) & \textbf{0.17 (0.06)} & 0.41 (0.10) & 0.06 (0.04) & 0.05 (0.03) & \textbf{38.40 (8.47)} & 0.15 (0.06) \\
    \bottomrule
    \end{tabular}
    \label{tab:acc}
\end{table*}

\subsubsection{External environment}
The external environment represents the typing interface, providing raw information about the current typing state so the agent can perceive context and make decisions. Based on the agent’s Gaze and Finger actions, the environment updates the input text, modifies available suggestions, and returns feedback depending on the gaze location.

To efficiently train the agent in suggestion use, we developed a custom suggestion system that can generate word suggestions and optionally simulate auto-correction. Existing open-source systems are either too slow (e.g., GPT-based language models~\cite{radford2019language}) for large-scale training or require substantial setup (e.g., Presage). Our system identifies words matching the current prefix and ranks them using a weighted score based on word length and frequency, derived from the first 5,000 entries of the Wiktionary:Frequency list\footnote{\url{https://en.wiktionary.org/wiki/Wiktionary:Frequency_lists/PG/2006/04/1-10000}}. By default, suggestions follow a weighted ranking similar to~\cite{quinn2016cost}, with adjusted weights to achieve accuracy comparable to commercial mobile keyboards~\cite{lehmann2023typing, 2025suggestion}. 
Because suggestions often operate alongside auto-correction, we also simulate this feature: if the agent stops typing and the word is within an edit distance below two from the target, the error is corrected automatically with 80\% accuracy~\cite{shi2024crtypist}. To validate the system, we extracted all words appearing at least three times in the WS-Gaze dataset~\cite{2025suggestion}, yielding about 1,080 distinct English words. The system provides correct suggestions for 65\% of these words (\textit{Accuracy}\footnote{Here, accuracy refers to the probability that a correct suggestion appears before a word is completed. Commercial systems typically range from 45\%–80\%, depending on context.}), with suggestions appearing after about 54\% of the word length, similar to commercial systems~\cite{2025suggestion}. The system is fast and highly tunable, allowing adjustments to suggestion accuracy, appearance timing, and other parameters.

\subsection{Implementation and training details}
\label{sec:training}
The training dataset consists of 1,080 unique words from the WS-Gaze dataset~\cite{2025suggestion}. In each episode, the environment randomly selects a word and a starting gaze position and initiates interaction with the agent. At each step, two potential suggestions are presented, similar to standard systems, allowing the agent to observe whether the correct suggestion is included. Note that the leftmost suggestion always mirrors the currently typed content, similar to standard mobile typing systems. 
We employ the Proximal Policy Optimization (PPO) algorithm~\cite{schulman2017proximal} with an LSTM-based architecture~\cite{hochreiter1997long} to capture temporal dependencies in sequential typing actions. Two-thumb and one-finger typing are simulated using different cognitive parameters, although the agent issues only one finger action per step.
We apply Bayesian Optimization~\cite{williamson2022bayesian} to fit the five cognitive parameters by minimizing the Jensen–Shannon divergence~\cite{lin2002divergence}, following the method used in previous work~\cite{shi2024crtypist, shi2025simulating}, and identify the best parameter set across all five parameters ($P_M$, $P_F$, $P_V$, $P_K$ and $P_S$) that aligns with the target metrics characterizing the corresponding human group behavior. More details are provided in Appendix~\ref{sec:train_details}.

\section{Model Evaluation}
\label{sec:results}

To better assess word suggestion models, we extend the benchmark proposed by~\citet{shi2024crtypist} with a set of new metrics characterizing typing behavior with suggestions, including \textit{Picked suggestions} (Picked; words for which the suggestion list was checked and a suggestion was selected), \textit{Failed suggestions} (Failed; words for which the suggestion list was checked but no suggestion was selected), \textit{Start checking character} (Start; percentage of characters typed before the first fixation on the suggestion list), \textit{Gaze ratio on the suggestion list} (Gaze Sugg), and \textit{Keystroke savings} (KS). In addition, we evaluate on the four main metrics from the previous benchmark: \textit{Gaze ratio on the keyboard} (Gaze kbd), \textit{Uncorrected error rate} (UER), \textit{Backspaces per character} (BPC), and \textit{Words per minute} (WPM). More details are provided in Appendix~\ref{sec:metrics}.

In the following, we use this benchmark to evaluate how well the WSTypist reproduces human behavior across typing groups and baseline systems, its ability to capture individual differences, and its generalization to settings without suggestions or auto-correction. We also investigate whether behavioral strategies observed in human typists also emerge in the simulation model, such as the effect of word length on suggestion usage. Our evaluation relies on two human datasets containing both keystroke and gaze data. The WS-Gaze dataset~\cite{2025suggestion} is the only available dataset of typing with word suggestions that includes gaze data, while the How-we-type-mobile dataset~\cite{jiang2020we} captures standard typing without suggestions and is used to assess generalizability beyond ITE settings. For each condition, except where specified, we trained eight agents and fitted their cognitive parameters to align as closely as possible with the target group’s behavior across the metrics. Each agent is then evaluated based on the average performance over 100 words (corresponding to ca. 20 sentences in a typical typing study~\cite{mackenzie2003phrase}) with reported values being means and standard deviations across all words and agents.

\paragraph{Accuracy on human-like behaviors}
As shown in~\autoref{tab:acc}, our model produces behavior that is similar to that of humans with respect to several important typing metrics. In particular, with respect to suggestion usage, the model replicates behavioral differences when comparing typing with two thumbs to typing with one finger. Both human participants and the model use word suggestions modestly in Two-thumb typing, resulting in limited gaze toward the suggestion list. In contrast, One-finger typing involves heavier reliance on suggestions and more frequent gazes toward the list, which also slows typing. Across all metrics, the agent remains within one standard deviation of human performance, capturing not only suggestion use but also the spatial distribution of gaze and error correction behavior. This consistency highlights the model’s ability to reproduce human decision-making and visual attention strategies in typing tasks on an aggregate level. In addition, we use the average (\textbf{Avg}) behavior across all users and conditions in the WS-Gaze dataset~\cite{2025suggestion} as a baseline for further evaluations.

\paragraph{Individual differences}
To evaluate the model's ability to replicate behavioral differences due to individuals' characteristics, such as typing speed, we selected two subgroups from the WS-Gaze dataset~\cite{2025suggestion}: the four fastest (H-WPM) and four slowest (L-WPM) typists. Results from both human participants and averaged agents indicate that the groups differ not only in typing speed but also in broader behavioral patterns. H-WPM users rely less on suggestions (lower Picked: 0.13 for human and 0.15 for the model) and rarely gaze at the suggestion list (lower Gaze Sugg: 0.17 for human and 0.13 for the model). L-WPM users, by contrast, depend heavily on suggestions and use backspace more often (higher BPC: 0.14 for human and 0.08 for the model). This might be a result of weaker typing skills or lower lexical proficiency, making the use of suggestions more beneficial (higher KS: 0.19 for human and 0.17 for the model) and also leading to better familiarity with the suggestion system (less Failed: 0.44 for human and 0.38 for the model). In both cases, the model reproduces human behavior across all metrics while capturing the differences between the groups. Full results are shown in~\autoref{tab:indi} Appendix~\ref{sec:eval_tables}.

\paragraph{Generalization to different ITE methods}
We tested the generalizability of our model to simulate human-like behavior for manual typing, and to simulate typing without auto-correction. 
To simulate manual typing, we slightly modified the RL model by removing suggestion-related actions (gaze on suggestions for Vision and pick for Finger). In addition, the suggestion list status in the observation space will always remain `Unknown'. We then retrained eight agents in the same way as described above and fitted the cognitive parameters to the behavioral metrics. We evaluate its performance by comparing the simulated behavior against human data from the How-we-type-mobile dataset~\cite{jiang2020we}, and against simulations produced by the CRTypist model~\cite{shi2024crtypist}. With orthographic processing incorporated, the simulation produces realistic and consistent results that are generally closer to human data than those of CRTypist. Full results are shown in~\autoref{tab:general_1} Appendix~\ref{sec:eval_tables}.
Next, we tested the model’s ability to simulate typing without auto-correction by disabling this function in the suggestion system. We focused on nine WS-Gaze dataset~\cite{2025suggestion} participants who reported not using auto-correction and compared their metrics with simulations under the same conditions. The results show that the model closely replicates their behavior. Interestingly, isolating this subgroup, compared to the 18 participants who used auto-correction in the empirical data, reveals patterns not discussed in the original paper~\cite{2025suggestion}: higher suggestion use (0.23 vs. 0.15), lower failure rates (0.33 vs. 0.41), and slightly higher backspace usage (0.09 vs. 0.08). This trend is intuitive: without auto-correction, users must either backspace to fix errors or rely more strategically on suggestions. The model reproduced these contrasts and generalized well across different ITE methods. Full results are shown in~\autoref{tab:general_2} Appendix~\ref{sec:eval_tables}.

\begin{figure}[h]
  \centering
    \includegraphics[width=0.68\linewidth]{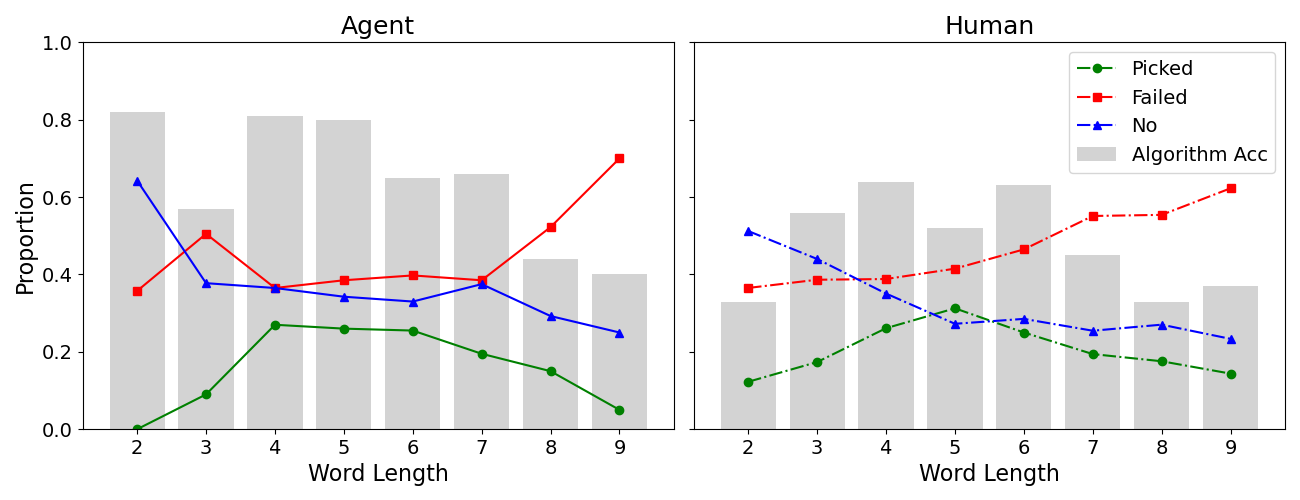}
    \caption{Effects of word length on suggestion usage by the agent, revealing a peak in selection rates for medium-length words and a decline for longer words. This trend is attributed to reduced algorithmic accuracy for longer suggestions and closely parallels patterns observed in human behavior.}
    \Description{This figure contrasts agent and human suggestion engagement across varying word lengths, highlighting how word length influences selection dynamics. The plotted curves show that both agents and humans exhibit optimal suggestion selection at mid-range word lengths, with performance decreasing as words grow longer. The graphs include the algorithm accuracy trace, offering context for the observed behavioral shifts. These visualizations underscore the interplay between system precision and cognitive effort, suggesting that longer words pose greater challenges for predictive systems and users alike.}
    \label{fig:acc_wl}
\end{figure}

\paragraph{Behavioral strategies}
We now present qualitative and quantitative results to examine whether specific human behavioral patterns of using word suggestions, identified by previous empirical works~\cite{lehmann2023typing, 2025suggestion}, also emerge in the simulations from the WSTypist agents. The results are drawn from the agents of the \textbf{Avg} group reported in \autoref{tab:acc}.
Specifically, the agents' typing behaviors demonstrate six empirically observed strategies~\cite{lehmann2023typing}: completing a word (Completion), correcting an error (Correction), directly selecting the next word (Prediction), adding apostrophes or capitals (Capitalization and Contraction), and choosing a different word before manually correcting it (Select and Modify). The frequency of each strategy is reported in \autoref{tab:strategies} in Appendix~\ref{sec:eval_tables}.
These strategies demonstrate the model’s flexibility and human-likeness. It adapts suggestion use to word context, mirroring how humans mix strategies to optimize efficiency and accuracy.
\autoref{fig:acc_wl} shows how the agent’s use of suggestions varies with target word length, which closely resembles a pattern observed from human behavior~\cite{2025suggestion}. Notably, suggestion usage peaks for words up to around six characters, after which usage declines while the rate of Failed suggestions increases. We attribute this to a drop in algorithmic accuracy beyond lengths of approximately six to seven characters, observed in both our and real-world suggestion systems. This pattern suggests that the agent develops an internal threshold for when suggestion use is most beneficial, yet still cannot anticipate failures originating from the suggestion algorithm.

\section{Using Simulations to Inform System Design}
\label{sec:app}

In this section, we demonstrate the applicability of our model to system design. Designing word suggestion systems or other ITE methods involves key challenges, such as balancing word length and frequency or prioritizing certain word types, whose effects often only emerge through time-consuming user studies. Evaluating new interfaces, such as inline suggestions or shortcut-based selection, is similarly costly. Typing is a highly practiced activity where users develop idiosyncratic strategies~\cite{feit2016, lehmann2023typing} and adapting to new designs can take hours and require many study sessions~\cite{mackenzie1999design, oulasvirta2013improving, zhang2021phraseflow, feit2014pianotext}. A/B testing in the wild is risky, as users may resist or abandon unfamiliar keyboards, while building new versions is itself expensive without clear evidence of benefit. Our simulation model offers an efficient alternative, enabling exploration of design trade-offs and rapid testing of prototypes. Crucially, it captures how users \textit{adapt} to new systems after deployment, which is an important consideration for typing applications used daily~\cite{palin2019people}.

We illustrate this through four design cases, each framed from the perspective of a developer improving a smartphone keyboard’s suggestion system. Based on observed user behavior (\textit{Observation}) from data logs or prior research, we formulate design hypotheses (\textit{Hypothesis}), prototype new variants, and use the WSTypist model to simulate behavior after adaptation (\textit{Simulation}). The results then inform a \textit{Recommendation} on whether and how to proceed. To simulate each system, we initialize four agents previously fitted to average user behavior in the WS-Gaze dataset (i.e., \textbf{Avg} in~\autoref{tab:acc}). We then continue training in the modified environment with a tenfold smaller learning rate, allowing gradual adaptation, and apply early stopping once behavior stabilizes. This setup approximates experienced users adapting to a new keyboard or suggestion algorithm over time. Across the four cases, we examine how suggestion accuracy affects behavior, whether prioritizing longer or capitalized words increases usage, and how interface changes, such as suggestion placement and shortcut-based selection, impact performance. Together, these examples highlight the model’s flexibility and its potential to inform the design of future suggestion systems.

\subsection{Accuracy-based system}
\label{sec:acc_based_sys}
\paragraph{Observation} Empirical data~\cite{2025suggestion} show that users experience varying success rates with word suggestions, with many instances where users check the suggestion list without selecting from it. It is unclear whether this comes from difficulty predicting when a word appears in the list or from personal strategies (e.g., using the list as a spell-check reference). A key design question is whether higher algorithm accuracy reduces failed checks and whether investing in more accurate algorithms is worthwhile.

\paragraph{Hypothesis} We hypothesize that suggestion accuracy strongly influences usage and that higher accuracy increases typing speed~\cite{delebecque:hal-02090402-condensed}.

\begin{figure}[h]
  \centering
  \begin{subfigure}[b]{0.38\linewidth}
    \centering
    \includegraphics[width=\linewidth]{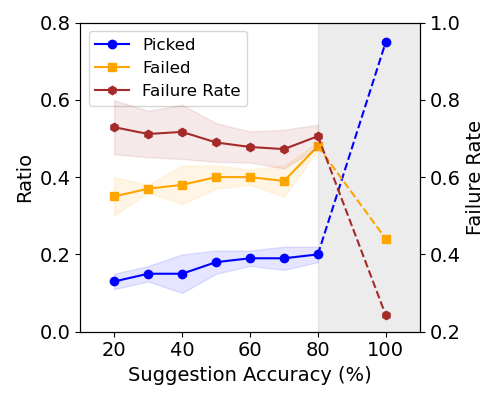}
    \caption{}
    \label{fig:acc_1}
  \end{subfigure}
  \begin{subfigure}[b]{0.38\linewidth}
    \centering
    \includegraphics[width=\linewidth]{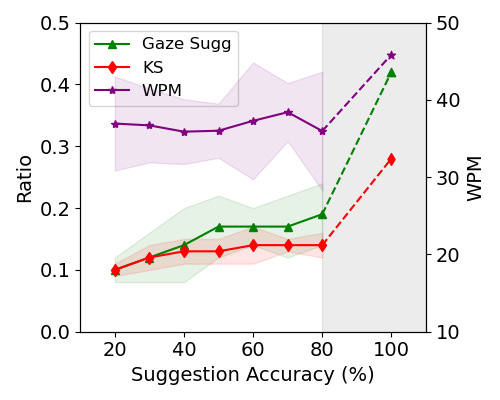}
    \caption{}
    \label{fig:acc_2}
  \end{subfigure}
  \caption{Influence of suggestion accuracy on user behavior, including effects on Picked, Failed suggestions, Keystroke Savings, Gaze ratio on suggestions, and typing speed (WPM) across different simulated accuracy levels. The shaded region highlights high-accuracy systems (>80\%), the dot marks oracle-level performance (100\%), and the dashed line projects a hypothetical trajectory toward optimal interaction.}
  \Description{This figure presents two plots illustrating how varying levels of suggestion accuracy shape user interaction patterns and typing efficiency. Panel (a) captures selection dynamics, showing a relationship between accuracy and failure rates, with Picked suggestions becoming more prevalent as suggestion accuracy improves. Panel (b) links perceptual and motor metrics, gaze behavior, keystroke savings, and typing speed. The shaded region and oracle marker contextualize the system’s operational thresholds, offering a benchmark for evaluating practical versus ideal outcomes.}
  \label{fig:acc_influence}
\end{figure}

\paragraph{Simulation} To test this, we simulated user behavior under systems with accuracy levels ranging from 20\% to 80\% (the system’s upper limit), in steps of about 10\%. To ensure fairness, all systems were configured so that the target word appeared in the suggestion list after roughly 54\% of its characters had been typed. \autoref{fig:acc_influence} shows the resulting agent behavior. As accuracy increased, the number of \emph{Picked} suggestions rose, accompanied by slight gains in keystroke savings and more gaze on the suggestion list. However, \emph{Failed} suggestions also increased, with a sharp jump at 80\%, suggesting agents began over-relying on the list and could not predict when a correct suggestion would appear. This led to a decrease in WPM. By contrast, simulating a system with perfect accuracy (100\%) yielded a very high \emph{Picked} rate, very low \emph{Failed} rate, and a clear speed increase.

\paragraph{Recommendation} Our findings suggest that \textbf{while higher accuracy encourages more suggestion use, it also raises the failure rate.} Gaze shifts without selection are costly, lowering typing speed and potentially increasing frustration. Designers should therefore aim for a balance: maintaining or improving accuracy while ensuring efficient use. \textbf{Accuracy in the 60–70\% range appears to offer the best trade-off.} Only if algorithm performance can be pushed well beyond the 80\% mark could it make sense to deploy a new algorithm to end-users.

\subsection{Length-priority system}
\paragraph{Observation} Typists were observed to choose fewer suggestions for shorter words, where keystroke savings are minimal, and word frequency is higher. Our agent replicated this behavior. As shown in~\autoref{fig:acc_wl}, words under five characters have low usage rates, even when suggestion accuracy is high.  

\paragraph{Hypothesis} We hypothesize that prioritizing longer words in the suggestion list can increase efficiency, as longer words are more costly to type manually and thus more beneficial to pick.

\begin{table}[h]
  \caption{Effects of prioritizing shorter vs. longer words in suggestions (Mean values).}
  \sffamily
  \footnotesize
  \begin{tabular}{lccccc}
    \toprule
    & Picked & Failed & Gaze Sugg & WPM & KS \\
    \midrule
    \textbf{Longer Priority} & 0.22 & 0.39 & 0.17 & 39.92 & 0.17 \\
    \textbf{Avg} & 0.19 & 0.41 & 0.17 & 38.40 & 0.15 \\
    \textbf{Shorter Priority} & 0.15 & 0.42 & 0.16 & 36.25 & 0.10 \\
    \bottomrule
    \end{tabular}
    \label{tab:longer}
\end{table}

\paragraph{Simulation} To test this, we configured the suggestion system so that ranking depended entirely on word length (ignoring frequency). We then compared two settings with nearly identical overall accuracy but different prioritization: one favoring shorter words, the other longer words. The results (\autoref{tab:longer}) show that agents benefited more when longer words were prioritized: they picked more suggestions (0.22 vs. 0.15), failed less often (0.39 vs. 0.42), saved more keystrokes (0.17 vs. 0.10), and typed faster (39.92 vs. 36.25). In contrast, prioritizing shorter words reduced utility: manual input was often faster than selecting a suggestion, so agents picked less.

\paragraph{Recommendation} Our findings suggest that \textbf{prioritizing longer words in the suggestion list can improve the efficiency and utility of suggestion-based input.} Future A/B testing in real-world systems may be warranted to validate these results.

\subsection{Capitalization-priority system}
\paragraph{Observation} Users differ in their accurate input of capitalized characters: some users intentionally skip capitalizing letters of proper nouns, relying on the suggestion list to correct them, while others manually capitalize them or omit capitalization entirely, accepting more errors in exchange for faster input~\cite{lehmann2023typing}. This might partially be driven by differences in typed text: users who type more proper nouns or type in languages with noun capitalization (e.g., German) might rely more on suggestions for capitalization.

\paragraph{Hypothesis} We hypothesize that users who frequently capitalize nouns learn to rely on suggestions for capitalization and will thus benefit from higher prioritization of capitalized suggestions. In contrast, reduced support for capitalized suggestions will discourage suggestion usage and hurt typing performance.

\begin{table}[h]
  \setlength{\tabcolsep}{3.7pt}
  \caption{Effects of prioritizing capitalized words in suggestions (Mean values).}
  \sffamily
  \footnotesize
  \begin{tabular}{lcccc}
    \toprule
    & Picked (Cap.) Sugg & Failed (Cap.) Sugg & Skip rate & WPM \\
    \midrule
    \textbf{High Priority} & 0.24 & 0.41 & 0.78 & 37.98 \\
    \textbf{Normal Priority} & 0.20 & 0.41 & 0.72 & 36.25 \\
    \textbf{Low Priority} & 0.18 & 0.38 & 0.62 & 37.13 \\
    \bottomrule
    \end{tabular}
    \label{tab:capital}
\end{table}

\paragraph{Simulation} To test this, we modified the model to include a binary indicator in the observation space signaling whether the next character should be capitalized. The “Type” action produces a capital letter when the indicator is active (doubling typing time), while the “NoAct” action allows the agent to skip capitalization. We then simulated three configurations of capitalization support in the suggestion system: \textbf{High Priority:} The capitalized form was presented as soon as it appeared within the top five suggestions; \textbf{Baseline:} The standard suggestion system used in previous experiments; \textbf{Low Priority:} 60\% of capitalized words that could have been suggested correctly were instead shown in lowercase. These configurations resulted in accuracies of 94\% (High), 58\% (Baseline), and 38\% (Low) for capitalized words. 
To simulate typing with frequent noun capitalization, we trained agents on a dataset containing 216 normal words and 187 capitalized words, raising the proportion of capitalized words to nearly half. Four agents were first trained from scratch on the Baseline system (with auto-correction disabled and stricter error tolerance) to establish a stable baseline. For the High and Low systems, we then fine-tuned these agents with a tenfold smaller learning rate, applying early stopping once performance stabilized. The results (\autoref{tab:capital}) show that agents exposed to the High Priority system select Capitalized suggestions more frequently than agents under alternative configurations (0.24 vs. 0.20 vs. 0.18). They also tend to skip typing initial capitals more often (0.78 vs. 0.72 vs. 0.62), anticipating correction via suggestions.

\paragraph{Recommendation} These findings suggest that capitalization handling should be personalized rather than fixed, as \textbf{emphasizing capitalized suggestions can benefit users who frequently type proper nouns or prefer capitalized input}. More broadly, users benefit from personalized support while also adapting their behavior to the underlying algorithm.

\subsection{Adaptation to new interfaces}
\paragraph{Observation} Users typically shift their gaze across three areas while typing: the keyboard, the suggestion list, and the input field. Each suggestion selection can require up to three times the effort of a normal keystroke, making the process costly. This suggests that interface design, specifically the placement of suggested words, plays a key role in shaping suggestion efficiency.  

\paragraph{Hypothesis} We hypothesize that placing the word suggestion directly in the input field and introducing a shortcut for selecting it from the keyboard will increase suggestion usage and improve typing efficiency.

\begin{table}[h]
  \setlength{\tabcolsep}{3.3pt}
  \caption{Performance comparison between list-based suggestions and alternative interfaces: inline suggestions (InputSugg) with or without Shortcut selection.}
  \sffamily
  \footnotesize
  \begin{tabular}{lcccccc}
    \toprule
    & Picked & Failed & Gaze Input & WPM & KS & Gaze shifts \\
    \midrule
    \textbf{Avg (List-based)} & 0.19 & 0.41 & 0.42 & 38.40 & 0.15 & 1.18 \\
    \textbf{InputSugg} & 0.24 & 0.36 & 0.63 & 42.32 & 0.18 & 0.74 \\
    \textbf{InputSugg + Shortcut} & 0.26 & 0.36 & 0.65 & 45.09 & 0.19 & 0.72 \\
    \bottomrule
    \end{tabular}
    \label{tab:interface}
\end{table}

\paragraph{Simulation} We simulated agents’ behavior using two modified systems to test this hypothesis. Gazing at the input field provides information about the suggestion status (i.e., whether the target word is suggested), while looking elsewhere leaves this information unknown. The action and state spaces were kept unchanged. In one system, we reduced the picking time to match a regular keystroke, simulating a keyboard shortcut for selecting the suggestion displayed in the input field. We compare agent behavior under the original system, a system with the top suggestion displayed in the input field, and a system with the additional keyboard shortcut. Agents in the new designs required fewer gaze shifts per word (decreasing from about 1.2 to 0.7), used suggestions more frequently (about 25\% on average), and allocated more gaze to the input field (about 64\%), resulting in higher typing speed. Interestingly, agents initially continued to look at the suggestion list even when unnecessary, but gradually learned to ignore it almost entirely. Full results are shown in~\autoref{tab:interface}.

\paragraph{Recommendation} Our simulations suggest that \textbf{placing the top suggestion directly in the input field can improve suggestion efficiency and typing performance.} As future work, we will build a prototype for A/B testing with users to validate these findings empirically.

\section{Discussion}
\label{sec:diss}

We presented and evaluated a computational model of typing behavior that simulates how humans type with word suggestions. The model serves as a tool for testing typing systems and gaining insights into cognitive behavior through the analysis of simulated agents. We discuss design implications and limitations below.

\subsection{Understanding diversity in human behavior}
\label{sec:under}
Empirical research shows that typing behavior with word suggestions varies substantially across users. Our simulations reveal similar variations in strategy, efficiency, and reliance on assistance features. For example, high-WPM users tend to minimize interaction with the suggestion list and rely on their typing fluency, whereas low-WPM users shift their gaze toward suggestions more frequently. However, typing speed is only one dimension of variation. Cognitive factors such as orthographic processes, working memory, and willingness to explore suggestions also play important roles.

By parameterizing these cognitive traits and preferences, our model captures these behavioral differences. This parameter space allows us to generate a wide spectrum of user profiles, from ``lazy'' typists who frequently check and select suggestions to expert typists who rarely divert attention away from the keyboard. Importantly, the model does not enforce a single \textit{optimal} interaction pattern but instead allows strategies to emerge from cognitive and motor constraints.
Although the preferences driving suggestion use and the reward structure governing action decisions are not fully understood, our ablation studies (Appendix~\ref{sec:abl}) indicate that incorporating $P_S$ into the reward function aligns with empirical behavior. Future work should further explore which cognitive and psychological factors influence typing behavior beyond speed and accuracy.

Modeling orthographic processing, which accounts for individual differences in linguistic knowledge, represents a meaningful step forward compared to previous typing models~\cite{shi2024crtypist, shi2025simulating}. Building on research in psycholinguistics~\cite{pinet2016typing, kandel2023written}, we extended the internal environment to include cognitive mechanisms related to language production that also influence typing behavior, which further increased the agents' similarity to human typing behavior in comparison to prior models. Thus, a next step for future work should be to also understand and model the higher-level processes involved in writing, that is, language planning, translation, and reviewing~\cite{hayes}, and how these affect typing.

\subsection{Using simulations to inform system design}
We consider the personalized typing systems that allow users to tailor their experience to their own needs. While personalized dictionaries are common in commercial keyboards, the hyperparameters of suggestion algorithms or UI designs are rarely personalized. For instance, some users may rely on suggestions for longer words or favor capitalized suggestions. Such customization could enhance efficiency by aligning system behavior with individual goals and habits, although it remains unclear how aware users are of their own preferences regarding suggestion use. Beyond manual customization, a more advanced approach involves systems that learn from users’ behavior over time. This shifts the interaction paradigm: rather than users adapting to a fixed system, the system adapts to users, fostering greater control and satisfaction. Future systems could be optimized for different user profiles. For example, by de-prioritizing Failed suggestions (using gaze data), supporting users with limited linguistic knowledge, or introducing gaze-based adaptive features, such as highlighting suggestions when fixated or offering personalized gaze-based shortcuts for selection. By continuously monitoring and analyzing typing patterns, the system could learn preferred words, common phrases, or stylistic tendencies and dynamically update suggestions.

Simulation models can play a key role in guiding such designs. With our model, agents can be parameterized with specific cognitive profiles while system parameters are varied, allowing us to identify which designs best support different behavioral patterns. For instance, our investigation of capitalization strategies was motivated by the idea that users who frequently type capitalized words might benefit from tailored support, whereas those who rarely use capitalization might not value such prioritization, or could even be hindered by it. More broadly, users with different cognitive profiles may require different forms of support. For example, individuals with strong linguistic knowledge might gain more from prioritizing longer words, as this maximizes keystroke savings, whereas those with weaker linguistic knowledge could benefit more from higher-frequency words that support spelling accuracy. Exploring such personalized adaptations offers a promising direction for future research.

\subsection{Limitations and future work}
\label{sec:limi}
Much of our evaluation (Section~\ref{sec:results}) uses the WS-Gaze dataset~\cite{2025suggestion}, the only typing-with-suggestions dataset that includes gaze data. Collected on participants’ own phones, it spans diverse displays and keyboards, allowing us to simulate varied user behaviors and adaptation. By tuning cognitive parameters and hyperparameters, we can reproduce the behavior of specific users or aggregate patterns across devices. Close alignment with human data also helps distinguish whether seemingly suboptimal behavior comes from algorithmic limitations, user strategies, or individual preferences.

Looking ahead, simulation tools should become more accessible to practitioners. Our current approach relies on expert tuning, but practical impact requires tools usable without deep reinforcement learning expertise. While simulations can accelerate early-stage design and hypothesis testing, they cannot replace user studies, which remain necessary to validate real-world experience. This is especially relevant for studying adaptation, where multi-session experiments are costly and slow. Simulation-based evaluations are an efficient means to speed up hypothesis testing and identify promising design candidates in early phases of design.

Future work could extend the Vision module to model reading more explicitly, for instance, through gaze-scanning mechanisms, enabling more realistic handling of longer or multi-item suggestions~\cite{buschek21}. Incorporating different error types may clarify how users balance correction strategies (e.g., backspacing vs. suggestions vs. auto-correction). Further directions include modeling linguistic knowledge and automaticity~\cite{logan2018automatic}, integrating intermediate linguistic processing (e.g., LLM-based models)~\cite{shi2025chartist, binz2023turning, nguyen2024predicting}, and accounting for psychological factors such as motivation, trust, and emotion~\cite{lee2014influence, lee2015influence, zhang2021motivation}. These, combined with hierarchical reinforcement learning~\cite{pateria2021hierarchical}, may improve realism, for example, by capturing frustration with prediction errors~\cite{alharbi2020effects, hertzum2023frustration} and its impact on suggestion use~\cite{toader2019effect}. As LLM-based writing tools~\cite{lin2024rambler, goodman2024lampost, buschek2024} become widespread, understanding user adaptation will be essential for designing systems that better align with human needs.

\section{Conclusion}

We present WSTypist, a cognitively informed reinforcement learning model that simulates human typing with word suggestions. The model captures key human-like behaviors, including gaze distribution, typing speed, and suggestion usage, and generalizes across different user groups and system settings. Across four use cases, we show how the simulation can inform the design of word suggestion algorithms and user interfaces. WSTypist combines fidelity to human performance with the flexibility to explore novel system configurations, making it a valuable tool for advancing our understanding of typing behavior and guiding the design of intelligent text entry systems.

\newpage

% ...

\bibliographystyle{ACM-Reference-Format}
\bibliography{reference}

@phdthesis{feit2018,
   author = {Anna Maria Feit},
   school = {Aalto University},
   pages = {182 + app. 56},
   publisher = {Aalto University},
   title = {Assignment Problems for Optimizing Text Input},
   url = {http://urn.fi/URN:ISBN:978-952-60-8016-1},
   year = {2018}
}

@article{hayes,
author = {John R. Hayes},
title ={Modeling and Remodeling Writing},

journal = {Written Communication},
volume = {29},
number = {3},
pages = {369-388},
year = {2012},
doi = {10.1177/0741088312451260},

URL = { 
    
        https://doi.org/10.1177/0741088312451260
    
    

},
eprint = { 
    
        https://doi.org/10.1177/0741088312451260
    
    

}
}

@inproceedings{buschek21,
author = {Buschek, Daniel and Z\"{u}rn, Martin and Eiband, Malin},
title = {The Impact of Multiple Parallel Phrase Suggestions on Email Input and Composition Behaviour of Native and Non-Native English Writers},
year = {2021},
isbn = {9781450380966},
publisher = {Association for Computing Machinery},
address = {New York, NY, USA},
url = {https://doi.org/10.1145/3411764.3445372},
doi = {10.1145/3411764.3445372},
booktitle = {Proceedings of the 2021 CHI Conference on Human Factors in Computing Systems},
articleno = {732},
numpages = {13},
location = {Yokohama, Japan},
series = {CHI '21}
}

@inproceedings{OPTI,
author = {MacKenzie, I. Scott and Zhang, Shawn X.},
title = {The design and evaluation of a high-performance soft keyboard},
year = {1999},
isbn = {0201485591},
publisher = {Association for Computing Machinery},
address = {New York, NY, USA},
url = {https://doi.org/10.1145/302979.302983},
doi = {10.1145/302979.302983},
booktitle = {Proceedings of the SIGCHI Conference on Human Factors in Computing Systems},
pages = {25–31},
numpages = {7},
location = {Pittsburgh, Pennsylvania, USA},
series = {CHI '99}
}

@inproceedings{feit2016,
  author = {Feit, Anna Maria and Weir, Daryl and Oulasvirta, Antti},
  title = {How We Type: Movement Strategies and Performance in Everyday Typing},
  year = {2016},
  isbn = {9781450333627},
  publisher = {Association for Computing Machinery},
  address = {New York, NY, USA},
  url = {https://doi.org/10.1145/2858036.2858233},
  doi = {10.1145/2858036.2858233},
  booktitle = {Proceedings of the 2016 CHI Conference on Human Factors in Computing Systems},
  pages = {4262–4273},
  numpages = {12},
  location = {San Jose, California, USA},
  series = {CHI '16}
}

@article{leino2024mobile,
  title={Mobile Typing with Intelligent Text Entry: A Large-Scale Dataset and Results},
  author={Leino, Katri and Laine, Markku and Kurimo, Mikko and Oulasvirta, Antti},
  journal={Preprint},
  year={2024},
  url={https://doi.org/10.21203/rs.3.rs-4654512/v1}
}

@article{alharbi2020effects,
  title={The effects of predictive features of mobile keyboards on text entry speed and errors},
  author={Alharbi, Ohoud and Stuerzlinger, Wolfgang and Putze, Felix},
  journal={Proceedings of the ACM on Human-Computer Interaction},
  volume={4},
  number={ISS},
  pages={1--16},
  year={2020},
  publisher={ACM New York, NY, USA},
  url={https://doi.org/10.1145/3427311}
}

@article{lehmann2023typing,
  title={Typing Behavior is About More than Speed: Users' Strategies for Choosing Word Suggestions Despite Slower Typing Rates},
  author={Lehmann, Florian and Kornecki, Itto and Buschek, Daniel and Feit, Anna Maria},
  journal={Proceedings of the ACM on Human-Computer Interaction},
  volume={7},
  number={MHCI},
  pages={1--26},
  year={2023},
  publisher={ACM New York, NY, USA},
  url={https://doi.org/10.1145/3604276}
}

@article{kristensson2009five,
  title={Five challenges for intelligent text entry methods},
  author={Kristensson, Per Ola},
  journal={AI Magazine},
  volume={30},
  number={4},
  pages={85--85},
  year={2009},
  url={https://doi.org/10.1609/aimag.v30i4.2269}
}

@article{kristensson2018statistical,
  title={Statistical Language Processing for Text Entry},
  author={Kristensson, Per Ola},
  journal={Computational Interaction},
  pages={43--64},
  year={2018},
  publisher={Oxford University Press Oxford, UK},
  url={https://doi.org/10.1093/oso/9780198799603.003.0003}
}

@article{hard2018federated,
  title={Federated learning for mobile keyboard prediction},
  author={Hard, Andrew and Rao, Kanishka and Mathews, Rajiv and Ramaswamy, Swaroop and Beaufays, Fran{\c{c}}oise and Augenstein, Sean and Eichner, Hubert and Kiddon, Chlo{\'e} and Ramage, Daniel},
  journal={arXiv preprint arXiv:1811.03604},
  year={2018},
  url={https://doi.org/10.48550/arXiv.1811.03604}
}

@article{chandramouli2024workflow,
  title={A Workflow for Building Computationally Rational Models of Human Behavior},
  author={Chandramouli, Suyog and Shi, Danqing and Putkonen, Aini and De Peuter, Sebastiaan and Zhang, Shanshan and Jokinen, Jussi and Howes, Andrew and Oulasvirta, Antti},
  journal={Computational Brain \& Behavior},
  pages={1--21},
  year={2024},
  publisher={Springer},
  url={https://doi.org/10.1007/s42113-024-00208-6}
}

@article{garay2006text,
  title={Text prediction systems: a survey},
  author={Garay-Vitoria, Nestor and Abascal, Julio},
  journal={Universal Access in the Information Society},
  volume={4},
  pages={188--203},
  year={2006},
  publisher={Springer},
  doi={10.1007/s10209-005-0005-9}
}

@inproceedings{fowler2015effects,
  title={Effects of language modeling and its personalization on touchscreen typing performance},
  author={Fowler, Andrew and Partridge, Kurt and Chelba, Ciprian and Bi, Xiaojun and Ouyang, Tom and Zhai, Shumin},
  booktitle={Proceedings of the 33rd annual ACM conference on human factors in computing systems},
  pages={649--658},
  year={2015},
  url={https://doi.org/10.1145/2702123.2702503}
}

@inproceedings{kristensson2021design,
  title={Design and analysis of intelligent text entry systems with function structure models and envelope analysis},
  author={Kristensson, Per Ola and M{\"u}llners, Thomas},
  booktitle={Proceedings of the 2021 CHI Conference on Human Factors in Computing Systems},
  pages={1--12},
  year={2021},
  url={https://doi.org/10.1145/3411764.3445566}
}

@article{hetzel2021,
  author = {Hetzel, Lorenz and Dudley, John and Feit, Anna Maria and Kristensson, Per Ola},
  title = {Complex Interaction as Emergent Behaviour: Simulating Mid-Air Virtual Keyboard Typing using Reinforcement Learning},
  year = {2021},
  issue_date = {2021},
  publisher = {IEEE Educational Activities Department},
  address = {USA},
  volume = {27},
  number = {11},
  issn = {1077-2626},
  url = {https://doi.org/10.1109/TVCG.2021.3106494},
  doi = {10.1109/TVCG.2021.3106494},
  journal = {IEEE Transactions on Visualization and Computer Graphics},
  pages = {4140–4149},
  numpages = {10}
}

@inproceedings{oulasvirta2022,
    author = {Oulasvirta, Antti and Jokinen, Jussi P. P. and Howes, Andrew},
    title = {Computational Rationality as a Theory of Interaction},
    year = {2022},
    isbn = {9781450391573},
    url = {https://doi.org/10.1145/3491102.3517739},
    doi = {10.1145/3491102.3517739},
    booktitle = {Proceedings of the 2022 CHI Conference on Human Factors in Computing Systems},
    articleno = {359},
    numpages = {14},
    series = {CHI '22}
}

@inproceedings{buschek2024,
    author = {Buschek, Daniel},
    title = {Collage is the New Writing: Exploring the Fragmentation of Text and User Interfaces in AI Tools},
    year = {2024},
    isbn = {9798400705830},
    url = {https://doi.org/10.1145/3643834.3660681},
    booktitle = {Proceedings of the 2024 ACM Designing Interactive Systems Conference},
    pages = {2719–2737},
    numpages = {19},
    series = {DIS '24}
}

@inproceedings{roy2021typing,
  title={Typing efficiency and suggestion accuracy influence the benefits and adoption of word suggestions},
  author={Roy, Quentin and Berlioux, S{\'e}bastien and Casiez, G{\'e}ry and Vogel, Daniel},
  booktitle={Proceedings of the 2021 CHI Conference on Human Factors in Computing Systems},
  pages={1--13},
  year={2021},
  url={https://doi.org/10.1145/3411764.3445725}
}

@inproceedings{yu2018device,
  title={On-device neural language model based word prediction},
  author={Yu, Seunghak and Kulkarni, Nilesh and Lee, Haejun and Kim, Jihie},
  booktitle={Proceedings of the 27th international conference on computational linguistics: system demonstrations},
  pages={128--131},
  year={2018},
  url={https://aclanthology.org/C18-2028}
}

@article{ghosh2017neuralnetworkstextcorrection,
  title={Neural networks for text correction and completion in keyboard decoding},
  author={Ghosh, Shaona and Kristensson, Per Ola},
  journal={arXiv preprint arXiv:1709.06429},
  year={2017},
  url={https://doi.org/10.48550/arXiv.1709.06429}
}

@inproceedings{kristensson2004,
  author = {Kristensson, Per-Ola and Zhai, Shumin},
  title = {SHARK2: a large vocabulary shorthand writing system for pen-based computers},
  year = {2004},
  isbn = {1581139578},
  url = {https://doi.org/10.1145/1029632.1029640},
  doi = {10.1145/1029632.1029640},
  booktitle = {Proceedings of the 17th Annual ACM Symposium on User Interface Software and Technology},
  pages = {43–52},
  numpages = {10},
  series = {UIST '04}
}

@inproceedings{quinn2016cost,
  title={A cost-benefit study of text entry suggestion interaction},
  author={Quinn, Philip and Zhai, Shumin},
  booktitle={Proceedings of the 2016 CHI conference on human factors in computing systems},
  pages={83--88},
  year={2016},
  url={https://doi.org/10.1145/2858036.2858305}
}

@inproceedings{jokinen2017modelling,
  title={Modelling learning of new keyboard layouts},
  author={Jokinen, Jussi PP and Sarcar, Sayan and Oulasvirta, Antti and Silpasuwanchai, Chaklam and Wang, Zhenxin and Ren, Xiangshi},
  booktitle={Proceedings of the 2017 CHI conference on human factors in computing systems},
  pages={4203--4215},
  year={2017},
  url={https://doi.org/10.1145/3025453.3025580}
}

@inproceedings{leiva2021we,
  title={How we swipe: A large-scale shape-writing dataset and empirical findings},
  author={Leiva, Luis A and Kim, Sunjun and Cui, Wenzhe and Bi, Xiaojun and Oulasvirta, Antti},
  booktitle={Proceedings of the 23rd International Conference on Mobile Human-Computer Interaction},
  pages={1--13},
  year={2021},
  url={https://doi.org/10.1145/3447526.3472059}
}

@inproceedings{banovic2019limits,
  title={The limits of expert text entry speed on mobile keyboards with autocorrect},
  author={Banovic, Nikola and Sethapakdi, Ticha and Hari, Yasasvi and Dey, Anind K and Mankoff, Jennifer},
  booktitle={Proceedings of the 21st International Conference on Human-Computer Interaction with Mobile Devices and Services},
  pages={1--12},
  year={2019},
  url={https://doi.org/10.1145/3338286.3340126}
}

@inproceedings{palin2019people,
  title={How do people type on mobile devices? Observations from a study with 37,000 volunteers},
  author={Palin, Kseniia and Feit, Anna Maria and Kim, Sunjun and Kristensson, Per Ola and Oulasvirta, Antti},
  booktitle={Proceedings of the 21st International Conference on Human-Computer Interaction with Mobile Devices and Services},
  pages={1--12},
  year={2019},
  url={https://doi.org/10.1145/3338286.3340120}
}

@inproceedings{jiang2020we,
  title={How we type: Eye and finger movement strategies in mobile typing},
  author={Jiang, Xinhui and Li, Yang and Jokinen, Jussi PP and Hirvola, Viet Ba and Oulasvirta, Antti and Ren, Xiangshi},
  booktitle={Proceedings of the 2020 CHI conference on human factors in computing systems},
  pages={1--14},
  year={2020},
  url={https://doi.org/10.1145/3313831.3376711}
}

@inproceedings{jokinen2021touchscreen,
  title={Touchscreen typing as optimal supervisory control},
  author={Jokinen, Jussi and Acharya, Aditya and Uzair, Mohammad and Jiang, Xinhui and Oulasvirta, Antti},
  booktitle={Proceedings of the 2021 CHI conference on human factors in computing systems},
  pages={1--14},
  year={2021},
  url={https://doi.org/10.1145/3411764.3445483}
}

@inproceedings{shi2024crtypist,
  title={CRTypist: Simulating Touchscreen Typing Behavior via Computational Rationality},
  author={Shi, Danqing and Zhu, Yujun and Jokinen, Jussi PP and Acharya, Aditya and Putkonen, Aini and Zhai, Shumin and Oulasvirta, Antti},
  booktitle={Proceedings of the CHI Conference on Human Factors in Computing Systems},
  pages={1--17},
  year={2024},
  url={https://doi.org/10.1145/3613904.3642918}
}

@article{toader2019effect,
  title={The effect of social presence and chatbot errors on trust},
  author={Toader, Diana-Cezara and Boca, Grațiela and Toader, Rita and M{\u{a}}celaru, Mara and Toader, Cezar and Ighian, Diana and R{\u{a}}dulescu, Adrian T},
  journal={Sustainability},
  volume={12},
  number={1},
  pages={256},
  year={2019},
  publisher={MDPI}
}

@article{2025suggestion,
  title={How We Type with Word Suggestions: Understanding Visual Attention and Checking Behavior during Mobile Text Input},
  author={Li, Yang and Feit, Anna Maria},
  journal={Proceedings of the ACM on Interactive, Mobile, Wearable and Ubiquitous Technologies},
  volume={9},
  number={3},
  pages={1--29},
  year={2025},
  publisher={ACM New York, NY, USA}
}

@inproceedings{shi2025simulating,
  title={Simulating Errors in Touchscreen Typing},
  author={Shi, Danqing and Zhu, Yujun and Fernandes Junior, Francisco Erivaldo and Zhai, Shumin and Oulasvirta, Antti},
  booktitle={Proceedings of the 2025 CHI Conference on Human Factors in Computing Systems},
  pages={1--13},
  year={2025}
}

@article{murray2022simulation,
  title={What simulation can do for HCI research},
  author={Murray-Smith, Roderick and Oulasvirta, Antti and Howes, Andrew and M{\"u}ller, J{\"o}rg and Ikkala, Aleksi and Bachinski, Miroslav and Fleig, Arthur and Fischer, Florian and Klar, Markus},
  journal={Interactions},
  volume={29},
  number={6},
  pages={48--53},
  year={2022},
  publisher={ACM New York, NY, USA}
}

@article{schulman2017proximal,
  title={Proximal policy optimization algorithms},
  author={Schulman, John and Wolski, Filip and Dhariwal, Prafulla and Radford, Alec and Klimov, Oleg},
  journal={arXiv preprint arXiv:1707.06347},
  year={2017}
}

@article{hochreiter1997long,
  title={Long short-term memory},
  author={Hochreiter, Sepp and Schmidhuber, J{\"u}rgen},
  journal={Neural computation},
  volume={9},
  number={8},
  pages={1735--1780},
  year={1997},
  publisher={MIT press}
}

@article{kaelbling1998planning,
  title={Planning and acting in partially observable stochastic domains},
  author={Kaelbling, Leslie Pack and Littman, Michael L and Cassandra, Anthony R},
  journal={Artificial intelligence},
  volume={101},
  number={1-2},
  pages={99--134},
  year={1998},
  publisher={Elsevier}
}

@incollection{hutton2019eye,
  title={Eye tracking methodology},
  author={Hutton, SB},
  booktitle={Eye movement research: An introduction to its scientific foundations and applications},
  pages={277--308},
  year={2019},
  publisher={Springer}
}

@inproceedings{chen2019gmail,
  title={Gmail smart compose: Real-time assisted writing},
  author={Chen, Mia Xu and Lee, Benjamin N and Bansal, Gagan and Cao, Yuan and Zhang, Shuyuan and Lu, Justin and Tsay, Jackie and Wang, Yinan and Dai, Andrew M and Chen, Zhifeng and others},
  booktitle={Proceedings of the 25th ACM SIGKDD international conference on knowledge discovery \& data mining},
  pages={2287--2295},
  year={2019}
}

@article{anderson2004integrated,
  title={An integrated theory of the mind.},
  author={Anderson, John R and Bothell, Daniel and Byrne, Michael D and Douglass, Scott and Lebiere, Christian and Qin, Yulin},
  journal={Psychological review},
  volume={111},
  number={4},
  pages={1036},
  year={2004},
  publisher={American Psychological Association}
}

@article{kieras1997overview,
  title={An overview of the EPIC architecture for cognition and performance with application to human-computer interaction},
  author={Kieras, Davis E and Meyer, Davis E},
  journal={Human--Computer Interaction},
  volume={12},
  number={4},
  pages={391--438},
  year={1997},
  publisher={Taylor \& Francis}
}

@article{salvucci2001integrated,
  title={An integrated model of eye movements and visual encoding},
  author={Salvucci, Dario D},
  journal={Cognitive Systems Research},
  volume={1},
  number={4},
  pages={201--220},
  year={2001},
  publisher={Elsevier}
}

@inproceedings{guiard2015mathematical,
  title={A mathematical description of the speed/accuracy trade-off of aimed movement},
  author={Guiard, Yves and Rioul, Olivier},
  booktitle={Proceedings of the 2015 British HCI Conference},
  pages={91--100},
  year={2015}
}

@article{lin2002divergence,
  title={Divergence measures based on the Shannon entropy},
  author={Lin, Jianhua},
  journal={IEEE Transactions on Information theory},
  volume={37},
  number={1},
  pages={145--151},
  year={2002},
  publisher={IEEE}
}

@inproceedings{bengio2009curriculum,
  title={Curriculum learning},
  author={Bengio, Yoshua and Louradour, J{\'e}r{\^o}me and Collobert, Ronan and Weston, Jason},
  booktitle={Proceedings of the 26th annual international conference on machine learning},
  pages={41--48},
  year={2009}
}

@article{baddeley2012working,
  title={Working memory: Theories, models, and controversies},
  author={Baddeley, Alan},
  journal={Annual review of psychology},
  volume={63},
  number={1},
  pages={1--29},
  year={2012},
  publisher={Annual Reviews}
}

@inproceedings{shi2025chartist,
  title={Chartist: Task-driven Eye Movement Control for Chart Reading},
  author={Shi, Danqing and Wang, Yao and Bai, Yunpeng and Bulling, Andreas and Oulasvirta, Antti},
  booktitle={Proceedings of the 2025 CHI Conference on Human Factors in Computing Systems},
  pages={1--14},
  year={2025}
}

@article{logan2018automatic,
  title={Automatic control: How experts act without thinking.},
  author={Logan, Gordon D},
  journal={Psychological Review},
  volume={125},
  number={4},
  pages={453},
  year={2018},
  publisher={American Psychological Association}
}

@article{binz2023turning,
  title={Turning large language models into cognitive models},
  author={Binz, Marcel and Schulz, Eric},
  journal={arXiv preprint arXiv:2306.03917},
  year={2023}
}

@inproceedings{nguyen2024predicting,
  title={Predicting and understanding human action decisions: Insights from large language models and cognitive instance-based learning},
  author={Nguyen, Thuy Ngoc and Jamale, Kasturi and Gonzalez, Cleotilde},
  booktitle={Proceedings of the AAAI Conference on Human Computation and Crowdsourcing},
  volume={12},
  pages={126--136},
  year={2024}
}

@article{lee2015influence,
  title={The influence of emotion on keyboard typing: An experimental study using auditory stimuli},
  author={Lee, Po-Ming and Tsui, Wei-Hsuan and Hsiao, Tzu-Chien},
  journal={PloS one},
  volume={10},
  number={6},
  pages={e0129056},
  year={2015},
  publisher={Public Library of Science San Francisco, CA USA}
}

@article{lee2014influence,
  title={The influence of emotion on keyboard typing: an experimental study using visual stimuli},
  author={Lee, Po-Ming and Tsui, Wei-Hsuan and Hsiao, Tzu-Chien},
  journal={Biomedical engineering online},
  volume={13},
  number={1},
  pages={81},
  year={2014},
  publisher={Springer}
}

@article{zhang2021motivation,
  title={Motivation, social emotion, and the acceptance of artificial intelligence virtual assistants—Trust-based mediating effects},
  author={Zhang, Shiying and Meng, Zixuan and Chen, Beibei and Yang, Xiu and Zhao, Xinran},
  journal={Frontiers in Psychology},
  volume={12},
  pages={728495},
  year={2021},
  publisher={Frontiers Media SA}
}

@article{goodman2024lampost,
  title={LaMPost: AI writing assistance for adults with dyslexia using large language models},
  author={Goodman, Steven M and Buehler, Erin and Clary, Patrick and Coenen, Andy and Donsbach, Aaron and Horne, Tiffanie N and Lahav, Michal and MacDonald, Robert and Michaels, Rain Breaw and Narayanan, Ajit and others},
  journal={Communications of the ACM},
  volume={67},
  number={9},
  pages={80--89},
  year={2024},
  publisher={ACM New York, NY, USA}
}

@inproceedings{lin2024rambler,
  title={Rambler: Supporting Writing With Speech via LLM-Assisted Gist Manipulation},
  author={Lin, Susan and Warner, Jeremy and Zamfirescu-Pereira, JD and Lee, Matthew G and Jain, Sauhard and Cai, Shanqing and Lertvittayakumjorn, Piyawat and Huang, Michael Xuelin and Zhai, Shumin and Hartmann, Bjoern and others},
  booktitle={Proceedings of the 2024 CHI Conference on Human Factors in Computing Systems},
  pages={1--19},
  year={2024}
}

@article{pateria2021hierarchical,
  title={Hierarchical reinforcement learning: A comprehensive survey},
  author={Pateria, Shubham and Subagdja, Budhitama and Tan, Ah-hwee and Quek, Chai},
  journal={ACM Computing Surveys (CSUR)},
  volume={54},
  number={5},
  pages={1--35},
  year={2021},
  publisher={ACM New York, NY, USA}
}

@book{sutton1998reinforcement,
  title={Reinforcement learning: An introduction},
  author={Sutton, Richard S and Barto, Andrew G and others},
  volume={1},
  number={1},
  year={1998},
  publisher={MIT press Cambridge}
}

@article{expectation-confirmation-model,
author = {Bhattacherjee, Anol},
title = {Understanding information systems continuance: an expectation-confirmation model},
year = {2001},
issue_date = {September 2001},
publisher = {Society for Information Management and The Management Information Systems Research Center},
address = {USA},
volume = {25},
number = {3},
issn = {0276-7783},
url = {https://doi.org/10.2307/3250921},
doi = {10.2307/3250921},
journal = {MIS Q.},
month = sep,
pages = {351–370},
numpages = {20}
}

@article{parasuraman2000,
author = {A. Parasuraman},
title ={Technology Readiness Index (Tri): A Multiple-Item Scale to Measure Readiness to Embrace New Technologies},

journal = {Journal of Service Research},
volume = {2},
number = {4},
pages = {307-320},
year = {2000},
doi = {10.1177/109467050024001},

URL = {   
        https://doi.org/10.1177/109467050024001
},
eprint = { 
        https://doi.org/10.1177/109467050024001
}
,

}

@article{hertzum2023frustration,
  author = {Hertzum, Morten and Hornb\ae{}k, Kasper},
  title = {Frustration: Still a Common User Experience},
  year = {2023},
  issue_date = {June 2023},
  publisher = {Association for Computing Machinery},
  address = {New York, NY, USA},
  volume = {30},
  number = {3},
  issn = {1073-0516},
  url = {https://doi.org/10.1145/3582432},
  doi = {10.1145/3582432},
  journal = {ACM Trans. Comput.-Hum. Interact.},
  month = jun,
  articleno = {42},
  numpages = {26}
}

@article{pinet2016typing,
  title={Typing is writing: Linguistic properties modulate typing execution},
  author={Pinet, Svetlana and Ziegler, Johannes C and Alario, F-Xavier},
  journal={Psychonomic bulletin \& review},
  volume={23},
  number={6},
  pages={1898--1906},
  year={2016},
  publisher={Springer}
}

@inproceedings{mackenzie1999design,
  title={The design and evaluation of a high-performance soft keyboard},
  author={MacKenzie, I Scott and Zhang, Shawn X},
  booktitle={Proceedings of the SIGCHI conference on Human Factors in Computing Systems},
  pages={25--31},
  year={1999}
}

@inproceedings{oulasvirta2013improving,
  title={Improving two-thumb text entry on touchscreen devices},
  author={Oulasvirta, Antti and Reichel, Anna and Li, Wenbin and Zhang, Yan and Bachynskyi, Myroslav and Vertanen, Keith and Kristensson, Per Ola},
  booktitle={Proceedings of the SIGCHI Conference on Human Factors in Computing Systems},
  pages={2765--2774},
  year={2013}
}

@inproceedings{zhang2021phraseflow,
  title={PhraseFlow: Designs and empirical studies of phrase-level input},
  author={Zhang, Mingrui Ray and Zhai, Shumin},
  booktitle={Proceedings of the 2021 CHI Conference on Human Factors in Computing Systems},
  pages={1--13},
  year={2021}
}

@inproceedings{feit2014pianotext,
  title={Pianotext: Redesigning the piano keyboard for text entry},
  author={Feit, Anna Maria and Oulasvirta, Antti},
  booktitle={Proceedings of the 2014 conference on Designing interactive systems},
  pages={1045--1054},
  year={2014}
}

@softwareversion{delebecque:hal-02090402-condensed,
  title = {Scilab},
  author = {Delebecque, François and Gomez, Claude and Goursat, Maurice and Nikoukhah, Ramine and Steer, Serge and Chancelier, Jean-Philippe},
  date = {1994-01},
  version = {1.1},
  hal_id = {hal-02090402},
  institution = {Inria},
  license = {Scilab license},
  repository = {https://github.com/scilab/scilab},
  url = {https://www.scilab.org/},
  swhid = {...}
}

@article{risko2016cognitive,
  title={Cognitive offloading},
  author={Risko, Evan F and Gilbert, Sam J},
  journal={Trends in cognitive sciences},
  volume={20},
  number={9},
  pages={676--688},
  year={2016},
  publisher={Elsevier}
}

@book{williamson2022bayesian,
  title={Bayesian Methods for Interaction and Design},
  author={Williamson, John H and Oulasvirta, Antti and Kristensson, Per Ola and Banovic, Nikola},
  year={2022},
  publisher={Cambridge University Press}
}

@article{radford2019language,
  title={Language models are unsupervised multitask learners},
  author={Radford, Alec and Wu, Jeffrey and Child, Rewon and Luan, David and Amodei, Dario and Sutskever, Ilya and others},
  journal={OpenAI blog},
  volume={1},
  number={8},
  pages={9},
  year={2019}
}

@article{darragh1990reactive,
  title={The reactive keyboard: A predictive typing aid},
  author={Darragh, John J and Witten, Ian H and James, Mark L.},
  journal={Computer},
  volume={23},
  number={11},
  pages={41--49},
  year={1990},
  publisher={IEEE}
}

@inproceedings{garay1997intelligent,
  title={Intelligent word-prediction to enhance text input rate (a syntactic analysis-based word-prediction aid for people with severe motor and speech disability)},
  author={Garay-Vitoria, Nestor and Gonzalez-Abascal, Julio},
  booktitle={Proceedings of the 2nd international conference on Intelligent user interfaces},
  pages={241--244},
  year={1997}
}

@article{koester1996effect,
  title={Effect of a word prediction feature on user performance},
  author={Koester, Heidi Horstmann and Levine, Simon},
  journal={Augmentative and alternative communication},
  volume={12},
  number={3},
  pages={155--168},
  year={1996},
  publisher={Taylor \& Francis}
}

@article{koester2002modeling,
  title={Modeling the speed of text entry with a word prediction interface},
  author={Koester, Heidi Horstmann and Levine, Simon P},
  journal={IEEE transactions on rehabilitation engineering},
  volume={2},
  number={3},
  pages={177--187},
  year={2002},
  publisher={IEEE}
}

@article{koester1997keystroke,
  title={Keystroke-level models for user performance with word prediction},
  author={Koester, Heidi Horstmann and Levine, Simon},
  journal={Augmentative and Alternative Communication},
  volume={13},
  number={4},
  pages={239--257},
  year={1997},
  publisher={Taylor \& Francis}
}

@article{koester1998model,
  title={Model simulations of user performance with word prediction},
  author={Koester, Heidi Horstmann and Levine, Simon},
  journal={Augmentative and Alternative Communication},
  volume={14},
  number={1},
  pages={25--36},
  year={1998},
  publisher={Taylor \& Francis}
}

@inproceedings{mackenzie2003phrase,
  title={Phrase sets for evaluating text entry techniques},
  author={MacKenzie, I Scott and Soukoreff, R William},
  booktitle={CHI'03 extended abstracts on Human factors in computing systems},
  pages={754--755},
  year={2003}
}

@article{anastaseni2025smartphones,
  title={What smartphones change about writing: The impact of word suggestions on orthographic processing},
  author={Anastaseni, Anna and Roy, Quentin and Perret, Cyril and Romano, Antonio and Kandel, Sonia},
  journal={Cognitive Neuropsychology},
  pages={1--26},
  year={2025},
  publisher={Taylor \& Francis}
}

@incollection{kandel2023written,
  title={Written production: The APOMI model of word writing: Anticipatory processing of orthographic and motor information},
  author={Kandel, Sonia},
  booktitle={Language production},
  pages={209--232},
  year={2023},
  publisher={Routledge}
}

@article{roy2025word,
  title={Are word suggestions beneficial? The effect of typing efficiency and suggestion accuracy},
  author={Roy, Quentin and Casiez, G{\'e}ry and Vogel, Daniel},
  journal={ACM Transactions on Computer-Human Interaction},
  year={2025},
  publisher={ACM New York, NY}
}

\appendix
    
\begin{table*}[t!]
  \caption{\textbf{WS-NoPs}, and \textbf{WS-NoCom} refer to versions of the full \textbf{WSTypist} model where the influence of $P_S$, or the awareness of \textit{Completeness}, respectively, is removed.}
  \sffamily
  \footnotesize
  \begin{tabular}{llllllllll}
    \toprule
    & Picked & Failed & Start & Gaze Sugg & Gaze Kbd & BPC & UER & WPM & KS \\
    \midrule
    \textbf{Human Avg} & 0.18 & 0.38 & 0.40 & 0.20 & 0.36 & 0.08 & 0.04 & 42.65 & 0.14 \\
    \textbf{WSTypist} & 0.19 & 0.41 & 0.35 & 0.17 & 0.41 & 0.06 & 0.05 & 38.40 & 0.15 \\
    \textbf{WS-NoPs}  & \textbf{0.07} & 0.20 & 0.31 & 0.08 & 0.68 & 0.04 & 0.06 & 39.29 & 0.05 \\
    \textbf{WS-NoCom} & 0.22 & \textbf{0.45} & \textbf{0.21} & \textbf{0.28} & 0.39 & 0.04 & 0.05 & 36.79 & 0.17 \\
    \bottomrule
    \end{tabular}
    \label{tab:abl}
\end{table*}

\newpage
\section{Quantitative evaluation metrics}
\label{sec:metrics}

We adapt four metrics from prior work~\cite{shi2024crtypist, feit2016, 2025suggestion} to evaluate basic typing performance, independent of word suggestion behavior:

\begin{itemize}
    \item \textbf{Words per minute (WPM):} Primary measure of typing speed, calculated as the number of characters in the final input, divided by the time spent (in minutes), assuming an average word length of five characters ~\cite{shi2024crtypist, feit2016}.
    \item \textbf{Gaze ratio on keyboard (Gaze Kbd):} Proportion of time spent looking at the keyboard, calculated by dividing total fixation duration on the keyboard by overall fixation duration ~\cite{shi2024crtypist, 2025suggestion}.
    \item \textbf{Backspaces per character (BPC):} A measure of editing behavior, calculated by dividing the total number of backspace presses by the number of characters in the final input~\cite{2025suggestion}. Note that this varies from the previous benchmark~\cite{shi2024crtypist}, which used the number of backspaces per sentence, which is sensitive to sentence length.
    \item \textbf{Uncorrected error rate (UER):} Main indicator of typing accuracy. Computed as the Levenshtein distance between final input and ground truth, divided by the length of the longer string ~\cite{shi2024crtypist, 2025suggestion}.
\end{itemize}

In addition, we introduce five new metrics that specifically capture word-suggestion–related performance:

\begin{itemize}
    \item \textbf{Picked suggestions (Picked):} Percentage of typed words for which the suggestion list was fixated and for which the user picked a suggestion~\cite{2025suggestion}.
    \item \textbf{Failed suggestions (Failed):} Percentage of typed words during which the suggestion list was fixated, but no suggestion was picked~\cite{2025suggestion}.
    \item \textbf{Start checking character (Start):} Percentage of typed characters within a word before the user fixates the suggestion list for the first time. Computed for each word individually, for which the suggestion list is fixated at least once~\cite{2025suggestion}.
    \item \textbf{Gaze ratio on suggestion list (Gaze Sugg):} Proportion of time spent looking at the suggestion list, calculated by dividing total fixation duration on the suggestion list by overall fixation duration across the interface~\cite{2025suggestion}.
    \item \textbf{Keystroke savings (KS):} Key metric for assessing the benefit of suggestion usage, calculated as 1 minus the ratio of total keystrokes to the number of characters in the final input~\cite{2025suggestion, lehmann2023typing}.
\end{itemize}

We regard these metrics as essential when evaluating typing models that simulate suggestion usage. 

\section{Ablation study}
\label{sec:abl}

In this section, we evaluate the influence of the newly introduced cognitive parameter, \textbf{Suggestion Reliance} ($P_S$), as well as the newly introduced Working memory component, \textbf{Completeness}. We conduct a thorough ablation study below, with all results averaged over four agents. \textbf{WS-NoPs}, and \textbf{WS-NoCom} refer to versions of the full \textbf{WSTypist} model where the influence of $P_S$, or the awareness of \textit{Completeness}, respectively, is removed. As shown in ~\autoref{tab:abl}, when all components are included, the agent gradually learns suggestion usage behavior in a manner similar to humans, serving as our full model (\textbf{WSTypist}). Without $P_S$ (\textbf{WS-NoPs}), the agent receives no direct reward for using suggestions and instead optimizes solely for speed and accuracy. Suggestion use can still emerge in this case, as efficient suggestion usage may reduce typing time and improve accuracy, thereby increasing overall reward. However, discovering this strategy is significantly harder (and sometimes impossible) without the explicit positive reward for $P_S$. This illustrates that users who are used to using suggestions or have prior experience with similar ITE features are more likely to quickly unlock the benefits of a new system and improve their efficiency. Without \textit{Completeness} (\textbf{WS-NoCom}), the agent is unaware of how much of the word has been typed. However, due to the influence of suggestion reliance and limited spelling knowledge, it allocates more gaze to the suggestion list and starts earlier, resulting in an inefficient but safe usage strategy. 



\section{More training details}
\label{sec:train_details}

An episode terminates when one of the following occurs: 60 steps (prevents infinite loops and ensures efficient training~\cite{sutton1998reinforcement}) have been taken, a suggestion is successfully picked, or the agent types one character beyond the target word length (representing a space input). Preliminary experiments showed most models converged to specific parameter ranges. To improve efficiency, we restricted each parameter to a narrower range while still permitting sufficient exploration: $P_M \in [0.0, 0.2]$, $P_F \in [0.25, 0.35]$, $P_K \in [0.05, 0.15]$, $P_S \in [0.5, 0.7]$, and $P_V \in [0.5, 0.7]$. 
Compared to previous models~\cite{shi2024crtypist, shi2025simulating}, we only simplify the Vision and Finger modules by retaining the same functions but replacing the trained neural models. Early in training, the agent often selected suggestions arbitrarily, picking words dissimilar to the target and earning no reward. To address this, we employed curriculum learning~\cite{bengio2009curriculum}. Initially, the agent can only select suggestions exceeding a similarity threshold, which is relaxed later. This encourages incremental exploration of suggestion strategies (e.g., Select and Modify~\cite{lehmann2023typing}) while avoiding premature convergence on suboptimal behaviors. 

\section{Tables for Evaluation}
\label{sec:eval_tables}
    
\begin{table*}[h]
  \setlength{\tabcolsep}{3.6pt}
  \caption{The model's ability to replicate individual differences between High- and Low-WPM user groups. We report the Mean (SD) value for each group. The modeling results show close alignment with human data across all metrics, with differences falling within one SD, demonstrating the model’s ability to capture even extreme human behaviors.}
  \sffamily
  \footnotesize
  \begin{tabular}{lllllllllll}
    \toprule
    && Picked & Failed & Start & Gaze Sugg & Gaze Kbd & BPC & UER & WPM & KS \\
    \midrule
    \textbf{H-WPM} & Human~\cite{2025suggestion} &  0.13 (0.10) & 0.34 (0.07) & 0.45 (0.04) & 0.17 (0.08) & 0.35 (0.12) & 0.05 (0.04) & 0.03 (0.02) & 59.16 (14.99) & 0.09 (0.09) \\
     & Model & 0.15 (0.07) & 0.38 (0.06) & 0.41 (0.04) & 0.13 (0.06) & 0.40 (0.09) & 0.08 (0.03) & 0.05 (0.02) & 48.26 (8.56) & 0.11 (0.06) \\
    \midrule
    \textbf{L-WPM} & Human~\cite{2025suggestion} &  0.28 (0.12) & 0.30 (0.04) & 0.44 (0.06) & 0.20 (0.11) & 0.35 (0.12) & 0.14 (0.06) & 0.04 (0.02) & 30.38 (5.89) & 0.19 (0.13) \\ 
     & Model & 0.24 (0.08) & 0.34 (0.06) & 0.38 (0.05) & 0.16 (0.06) & 0.42 (0.07) & 0.08 (0.04) & 0.03 (0.02) & 33.49 (7.15) & 0.17 (0.08) \\
    \bottomrule
    \end{tabular}
    \label{tab:indi}
\end{table*}

\label{sec:gen}
\begin{table*}[h]
  \caption{Model generalizability across baseline systems without ITEs, compared with CRTypist~\cite{shi2024crtypist}. We report the Mean (SD) value for each group. Our model shows closer alignment with human data across all metrics compared to CRTypist.}
  \sffamily
  \footnotesize
  \begin{tabular}{llllll}
    \toprule
    \textbf{ITE} & & Gaze Kbd & BPC & UER & WPM \\
    \midrule
    \textbf{None} & Human~\cite{jiang2020we} & 0.60 (0.16) & 0.18 (0.14) & <0.01 (0.01) & 39.30 (10.3) \\
     & Model & 0.66 (0.09) & 0.13 (0.08) & <0.01 (0.01) & 35.41 (7.23) \\
     & CRTypist~\cite{shi2024crtypist} & 0.73 (0.04) & 0.26 (0.21) & <0.01 (0.01) & 34.80 (6.20) \\
    \bottomrule
    \end{tabular}
    \label{tab:general_1}
\end{table*}

\begin{table*}[h]
  \setlength{\tabcolsep}{3.7pt}
  \caption{Model generalizability across Word Suggestions (WS) and Word Suggestions with Auto-Correction (WS+AC). We report the Mean (SD) value for each group. The modeling results closely align with human data across all metrics, with differences within one SD, demonstrating the model’s ability to capture various ITE supports.}
  \sffamily
  \footnotesize
  \begin{tabular}{lllllllllll}
    \toprule
    \textbf{ITE} && Picked & Failed & Start & Gaze Sugg & Gaze Kbd & BPC & UER & WPM & KS \\
    \midrule
    \textbf{WS} & Human~\cite{2025suggestion} & 0.23 (0.10) & 0.33 (0.09) & 0.43 (0.07) & 0.19 (0.09) & 0.40 (0.13) & 0.09 (0.05) & 0.04 (0.02) & 39.55 (10.84) & 0.18 (0.08) \\ 
     & Model & 0.22 (0.07) & 0.36 (0.05) & 0.37 (0.06) & 0.16 (0.07) & 0.43 (0.09) & 0.07 (0.03) & 0.05 (0.02) & 34.94 (7.68) & 0.17 (0.06) \\
    \midrule
    \textbf{WS+AC} & Human~\cite{2025suggestion} & 0.15 (0.11) & 0.41 (0.06) & 0.37 (0.07) & 0.21 (0.10) & 0.34 (0.14) & 0.08 (0.05) & 0.04 (0.04) & 44.26 (13.76) & 0.14 (0.12) \\ 
     & Model & 0.15 (0.08) & 0.39 (0.07) & 0.36 (0.06) & 0.16 (0.06) & 0.39 (0.07) & 0.07 (0.04) & 0.04 (0.03) & 38.36 (8.83) & 0.12 (0.07) \\
    \bottomrule
    \end{tabular}
    \label{tab:general_2}
\end{table*}

\begin{table*}[h]
    \caption{Frequency of six word suggestion strategies in model behavior. Values are averaged across eight agents (Avg group) and reported as the mean percentage of words using suggestions for each strategy over 200-word trials per agent.}
    \label{tab:strategies}
    \centering
    \sffamily
    \footnotesize
    \setlength{\tabcolsep}{6pt}
    \begin{tabular}{p{2.8cm} p{10.2cm} p{1.5cm}}
    \toprule
    \textbf{Strategy} & \textbf{Description} & \textbf{Frequency} \\
    \midrule
    Completion & Type the beginning of a word and select a suggestion to complete it (e.g., ``prob'' $\rightarrow$ ``problem''). & 57.04\% \\
    Correction & Type a word with an error and select a suggestion to correct it (e.g., ``methad'' $\rightarrow$ ``method''). & 23.36\% \\
    Prediction & Select a suggestion before typing any characters. & 13.86\% \\
    Contraction & Type a contraction without punctuation and select a suggestion to complete it (e.g., ``dont'' $\rightarrow$ ``don’t''). & 1.04\% \\
    Capitalization & Type a word in lowercase and select a suggestion to apply capitalization (e.g., ``friday'' $\rightarrow$ ``Friday''). & 1.39\% \\
    Select-and-modify & Select a suggestion and then modify it (e.g., ``long'' $\rightarrow$ ``longer''). & 3.30\% \\
    \bottomrule
    \end{tabular}
\end{table*}

\end{document}